\documentclass[11pt,a4paper]{article}
\pdfoutput=1
\usepackage{jheppub}

\usepackage{multirow}
\usepackage{natbib}
\usepackage{amsmath,amssymb,amsfonts}
\usepackage{graphicx}
\usepackage{color}
\usepackage{bbm}
\usepackage{url}
\usepackage{slashed}
\usepackage{subfigure}
\usepackage[usenames,dvipsnames]{xcolor}
\usepackage{float} 
\usepackage{epsfig}
\usepackage{graphicx}
\usepackage{euscript}
\usepackage{color}
\usepackage{slashed}
\usepackage{epstopdf}
\usepackage{mathbbol}
\usepackage{mathrsfs}
\usepackage[normalem]{ulem}
\usepackage{pifont}
\usepackage{bbold}
\usepackage{float}
\allowdisplaybreaks
\usepackage{bbding}

\usepackage{verbatim}
\usepackage{graphicx}
\usepackage{latexsym}

\newcommand{\Eqref}[1]{Eq.~\eqref{#1}}

\newcommand{\be}{\begin{equation}}
\newcommand{\ee}{\end{equation}}
\def \eea{\end{eqnarray}}
\def \bea{\begin{eqnarray}}

\newcommand{\p}{\partial}

\definecolor{darkgreen}{rgb}{0.2,0.6,0}
\definecolor{lightblue}{rgb}{0,0.5,0.8}
\definecolor{lightred}{rgb}{0.8,0.2,0.2}
\definecolor{darkorange}{rgb}{1,0.549,0}
\definecolor{brown}{rgb}{0.609, 0.164, 0.164}

\allowdisplaybreaks

\title{
Starobinsky-inflation in asymptotically safe shift-symmetric scalar-tensor theory	
}
\author[a,b]{Benjamin Koch,} \emailAdd{benjamin.koch@tuwien.ac.at}
\affiliation[a]{ Institut fur Theoretische Physik, Technische Universitat Wien,
Wiedner Hauptstrasse 8-10, A-1040 Vienna, Austria
}
\affiliation[b]{Instituto de F\'isica, Pontificia Universidad Cat\'olica de Chile, Casilla 306, Santiago, Chile
}

\author[c]{Cristobal Laporte,} \emailAdd{calaporte@icc.ub.edu}
\affiliation[c]{Departament de Física Quàntica i Astrofísica, Institut de Ciències del Cosmos, Universitat de Barcelona, Martí i Franquès 1, E-08028 Barcelona, Spain}
\affiliation[d]{Institute for Mathematics, Astrophysics and Particle Physics (IMAPP),
Radboud University, Heyendaalseweg 135, 6525 AJ Nijmegen, The Netherlands}

\author[d]{Frank Saueressig,} \emailAdd{f.saueressig@science.ru.nl}

\abstract{
We investigate the constraining power of scalaron-driven inflation on asymptotically safe scalar-tensor theories. Starting from a Horndeski-type theory and applying a renormalization group improvement procedure generates higher-derivative couplings which are fixed in terms of the microscopic parameters – a structure that is expected to occur also within first principle computations based on the asymptotic safety mechanism. The latter are taken to be the free parameters appearing at the Gaussian fixed point. We find that the free parameter initially associated with the non-minimal gravity-matter coupling is not confined to the gravity-matter sector of the theory and also enters the effective higher-derivative couplings in the gravitational sector. 
We review the setting of multi-field inflationary models which is appropriate to analyze the inflationary dynamics in this context and illustrate their applicability by working out the explicit bounds on the non-minimal gravity-matter coupling resulting from cosmological observations. Given the fixed point structure of asymptotically safe scalar-tensor theories, the results indicate that UV-completions by two of the three available non-Gaussian fixed points can be ruled out while pinpointing phenomenologically viable RG trajectories emanating from the third fixed point.
}

\begin{document}
\maketitle
\section{Introduction}
\label{sect.Intro}
Scalar-tensor models provide a rich laboratory for testing new ideas about fundamental physics and developing phenomenological models which can be confronted with observations.  The framework supplements the gravitational degrees of freedom with a scalar field which may be interpreted as a fifth force. The consequences of this addition have been explored in a vast range of models in the realm of black holes and cosmology. In particular the phenomenology associated with Horndeski-type theories  \cite{Horndeski:1974wa,Nicolis:2008in,Deffayet:2009wt,Kobayashi:2019hrl} has been studied in much detail. In this case the admissible interactions are restricted by the requirement that the equations of motion are second order and do not give rise to higher-order derivative terms. In terms of black hole physics, this setting  allows to bypass the no-hair theorems known from general relativity \cite{Hui:2012qt,Sotiriou:2013qea,Kobayashi:2014eva,Babichev:2015rva,Babichev:2016fbg,Babichev:2016rlq,Ogawa:2015pea,Babichev:2017guv,Benkel:2016rlz,Hajian:2020dcq,Khoury:2020aya,Creminelli:2020lxn}. In terms of cosmology, Horndeski-type theories have recently been evoked to alleviate the Hubble tension arising from the measurement of the expansion of the universe at early and late times \cite{Barreira:2014jha,Renk:2017rzu,Frusciante:2019puu,Zumalacarregui:2020cjh,Petronikolou:2021shp}.

From the perspective of quantum gravity, scalar-tensor models constitute an important stepping stone when generalizing ideas developed in the context of pure gravity to the realm of more realistic gravity-matter systems. In particular, the gravitational asymptotic safety program \cite{Percacci:2017fkn,Reuter:2019byg,Reuter:2012id,Eichhorn:2018yfc,Pawlowski:2020qer,Saueressig:2023irs} has recently devoted significant attention to exploring the renormalization group flows of these models at the non-perturbative level. As key results, it has been shown that the non-Gaussian fixed point (called the Reuter fixed point) providing the high-energy completion of pure gravity within the asymptotic safety program, has a counterpart with very similar properties when the scalar degree of freedom is included. Quite remarkably, a minimal coupling of matter to gravity is not self-consistent in these settings \cite{Eichhorn:2012va}. The interactions with gravity at the quantum level induce non-trivial matter-selfinteractions as well as non-minimal couplings build from the spacetime curvature and the matter fields. The structure of these terms is not arbitrary though. In \cite{Laporte:2021kyp} it has been shown that the renormalization group flow of these models closes on the subspace of interaction monomials which are compatible with the shift-symmetry of the scalar kinetic term. This entails that scalar potentials are not generated by integrating out quantum fluctuations a priori.\footnote{For a recent discussion on generating a non-trivial Higgs potential within asymptotically safe gravity-matter systems, see \cite{Pastor-Gutierrez:2022nki}. In the context of cosmology, the generation of a non-trivial scalar potential by adding so-called kinetials in the approximation of the effective average action has been discussed in \cite{Percacci:2015wwa,Maitiniyazi:2025pou}, also see \cite{Wetterich:2022ncl} for review and more references.}
This raises the intriguing question of whether one can still obtain interesting cosmological models within the subspace of shift-symmetric scalar-tensor theories.

The goal of our work is then two-fold. First, we investigate which inflationary scenarios are compatible with the shift-symmetric scalar-tensor theories. Along these lines, we conclude that Starobinski-type models appear as natural candidates for inflation within this setting. We then explore how these scenarios can be used to constrain the free parameters appearing in the low-energy effective action.

Our approach follows the idea of renormalization group improvements \cite{Bonanno:2000ep,Bonanno:2001xi,Bonanno:2001hi} prominent within phenomenological studies within the gravitational asymptotic safety program in the context of black hole physics \cite{Reuter:2006rg,Koch:2013owa,Koch:2014cqa,Koch:2015nva,Platania:2019kyx,Knorr:2022kqp,Platania:2023srt}, galaxy rotation curves \cite{Reuter:2004nx,Reuter:2004nv}, the gravitational collapse \cite{Bonanno:2016dyv,Bonanno:2017zen,Bonanno:2023rzk}
, and cosmology \cite{Bentivegna:2003rr,Reuter:2003ca,Reuter:2005kb,Bonanno:2010mk,Bonanno:2018gck,Hindmarsh:2011hx,Zarikas:2024chv} and its generalization to the scale-setting program \cite{Contreras:2017eza,Hernandez-Arboleda:2018qdo,Contreras:2019cmf,Alvarez:2020xmk,
Alvarez:2022wef,Koch:2025gaw,Neckam:2025kip}.\footnote{For the construction of asymptotically safe cosmologies going beyond the framework of RG-improvements, see \cite{Gubitosi:2018gsl,Gubitosi:2019ncx}.}
 The incentive of these programs is to incorporate the ``leading quantum corrections in the analysis'' without the need of computing the corresponding couplings in the effective action. Typically, these studies construct the dynamics utilizing the scaling of the theory in the vicinity of the non-Gaussian fixed point providing the high-energy completion.  In contrast to that, we base our analysis on a renormalization group improvement at the Gaussian fixed point which controls the dynamics below the Planck scale. Since inflation is supposed to occur below the Planck scale, this strategy for incorporating quantum corrections seems to be better justified.

As our key result, we conclude that the effective dynamics obtained from the renormalization group improvement supports Starobinsky-type inflation models with the scalar field providing an additional flat direction in the inflationary potential. The closure of the renormalization group flow on the space of shift-symmetric interactions then implies that this flat direction is protected against quantum corrections. Demanding that the theory is compatible with the fluctuation spectra observed in the cosmic microwave background (CMB) then allows to constrain the free parameters of the theory. This is demonstrated explicitly at the level of the coupling associated with $RX$-interactions with $R$ being the Ricci scalar and $X$ the scalar kinetic term. 

The remainder of our work is organized as follows. Sect.\ \ref{Sect:OverviewFRG} reviews the basic concepts of Wilsonian renormalization group flows and Asymptotic Safety. Subsequently, Sect.\ \ref{sect.3} introduces the idea of renormalization group improvements and compares the strategies of ``improving at the Gaussian fixed point'' and the traditional path of ``improving at the non-Gaussian fixed point''. In particular, the renormalization group improved action underlying our analysis is given in Eq.\ \eqref{IRRGimprovedLagrangian}. In Sect.\ \ref{sect.4}, we convert this action into the Einstein-frame and the formalism for investigating inflation in the resulting setting is detailed in Sect.\ \ref{sect.5}. The results for the spectral indices are obtained in Sect.\ \ref{sect.6} and we use them to put bounds on the free parameters of the RG-improved model in Sect.\ \ref{sect.7}. Our conclusions are presented in Sect.\ \ref{sect.conclusions}.

\section{Asymptotically safe scalar-tensor models}
\label{Sect:OverviewFRG}
For our purpose, it is convenient to adopt a Wilsonian perspective on quantum field theory. The key idea of this approach is to organize the quantum fluctuations of a system in terms of their momentum $p^2$ and integrate out fluctuations with momenta larger than the coarse-graining scale $k$. Lowering $k$ corresponds to integrating out further shells of fluctuations. This induces a scale-dependence in the couplings, the Wilsonian renormalization group (RG) flow of the theory. 
This scale-dependence can be computed by employing functional RG techniques \cite{Wetterich:1992yh,Morris:1993qb,Reuter:1996cp}. Solving the flow equation generates a one-parameter family of effective average actions $\Gamma_k$ which include the quantum corrections with $p^2 \gtrsim k^2$ and provide an effective description of the physics at scales $p^2 \approx k^2$. 

The information about the running of the couplings $g_i(k)$ is  stored in the beta functions of the theory,
\be\label{betadef}
k \p_k g_i(k) = \beta_i(\{g_j\}) \, , 
\ee
written in terms of the dimensionless couplings $\{g_j\}$. The solutions of this coupled set of autonomous, first order equations constitute the RG trajectories of the theory. An essential feature of the beta functions are their fixed points $\{g_j^*\}$ where, by definition, $\beta_i(\{g_j^*\}) = 0, \forall i$. The RG flow in the vicinity of a fixed point can be studied by linearizing the beta functions,
\begin{equation}\label{linearizedBetaFunctions}
    \beta_i = \sum^{\infty}_{j=0} B_i{}^j \, \left(g_j - g^*_j\right) + O\left(\left(g_j - g^*_j\right)^2\right) \, . 
\end{equation}
Here $B_i{}^j \equiv \left. \frac{\partial \, \beta_i}{\partial \, g_j}\right|_{g=g^*}$ is the stability matrix whose eigenvalues and right-eigenvectors are denoted by $\lambda_J \equiv -\theta_J$ and $e_J$, respectively. The index $J$ enumerates the eigenvalues. Since the stability matrix $B$ is real but not necessarily symmetric, some eigenvalues could appear as complex conjugate pairs. 

The linearized flow equation can then be solved analytically 
\begin{equation}\label{gSolution}
    g_i(k) = g^{*}_i + \sum_{J} c_{J} \, e_i{}^J \, (k/k_0)^{-\theta_J} \, . 
\end{equation}
The $c_J$ are integration constants specifying the solution and $k_0$ is an arbitrary reference scale. This equation encodes the condition of an RG trajectory approaching $g_i^*$ in the limit $k \rightarrow \infty$. This requires that the coefficients $c_J$ associated with the stability coefficients coming with Re$(\theta_J) < 0$ must be set to zero. Conversely, the coefficients $c_J$ related to eigendirections where Re$(\theta_J) > 0$ are free parameters. Stipulating that there are only finitely many UV-attractive directions, the fixed point leads to a predictive theory with a finite number of free parameters. 

If the $\theta_J$ agree with the canonical mass-dimension of the dimensionful couplings, the underlying fixed point is called a Gaussian fixed point (GFP), since it corresponds to a free theory. There can also be non-Gaussian fixed points (NGFPs) where the stability coefficients receive quantum corrections.
If the RG-flow of a theory approaches a fixed point at high energy, it is expected that the theory is free from unphysical divergences \cite{Percacci:2017fkn}. The core idea of asymptotically safe scalar-tensor models is then that the UV-completion of the model is provided by a NGFP. 

The focus of our work is on asymptotically safe scalar-tensor theories whose field content is given by gravity supplemented by one real scalar field. The renormalization group flow of this class of models has been studied in \cite{Griguolo:1995db,Shaposhnikov:2009pv,Narain:2009fy,Narain:2009gb,Zanusso:2009bs,Eichhorn:2012va,Dona:2013qba,Labus:2015ska,Meibohm:2015twa,Dona:2015tnf,Percacci:2015wwa,Hamada:2017rvn,Eichhorn:2017sok,Eichhorn:2017als,Biemans:2017zca,Alkofer:2018fxj,Alkofer:2018baq,Pawlowski:2018ixd,Wetterich:2019rsn,Eichhorn:2020kca,Eichhorn:2020sbo,Eichhorn:2021tsx,deBrito:2021pyi,Knorr:2022ilz}. This has led to several structural insights. Firstly, the Reuter fixed point \cite{Reuter:1996cp} found in the case of pure gravity has an analogue once the scalar field is added \cite{Percacci:2003jz,Dona:2013qba,Biemans:2017zca,Laporte:2022ziz,deBrito:2023myf}. Secondly, a minimal coupling of the scalar field to gravity is not self-consistent if the theory is asymptotically safe \cite{Eichhorn:2012va}. Graviton fluctuations induce non-minimal matter self-interactions as well as non-minimal gravity-matter couplings. Thirdly, it was proven in \cite{Laporte:2021kyp} that the shift-symmetry of the scalar kinetic term is preserved under the RG flow. This severely restricts the type of non-minimal interaction terms which are generated by quantum effects. Lastly, \cite{Laporte:2021kyp} performed an elaborate scan of the fixed point structure on the space of (Euclidean signature) scalar-tensor theories, tracking the scale-dependence of the ansatz
\be\label{eq.ansmaster}
\Gamma_k^{\rm}[\phi,g] \simeq \int \mathrm{d}^4x \sqrt{g} \left[ \frac{2 \Lambda_k - R}{16 \pi \, G_k} + \frac{Z_k}{2} X
+ Z_k^2 \, C_k \, X^2 + Z_k \tilde{C}_k R^{\mu\nu} X_{\mu\nu} + Z_k \, D_k \, R \, X
\right] \, , 
\ee
supplemented by suitable gauge-fixing and ghost terms. Here $X \equiv g^{\mu\nu} (\partial_\mu \phi) \, (\partial_\nu \phi)$, and the scale-dependent couplings retained by the ansatz comprise the dimensionful Newton's coupling $G_k$, the cosmological constant $\Lambda_k$, a wave-function renormalization for the scalar field $Z_k$ as well as the non-minimal scalar-self interaction $C_k$ and couplings of the scalars to the spacetime curvature $\tilde{C}_k$ and $D_k$. Analyzing this system together with its subsystems obtained by setting some of the non-minimal couplings to zero, it was found that the corresponding beta functions always admit a GFP. In addition, there is a rich set of NGFPs. In particular, the inclusion of the coupling $D_k$ plays a crucial role in obtaining a fixed point structure suitable for rendering the approximation stable. 

Phenomenologically interesting solutions of the RG equation then emanate from a NGFP as $k \rightarrow \infty$. This equips the construction with predictive power. In order to exhibit a ``classical'' phase at low energy, the RG flow has to undergo a crossover to the GFP. In this regime, the dimensionful couplings freeze out and take their IR values $\{\Lambda_0, G_0, C_0, \tilde{C}_0, D_0\}$. These couplings are determined in terms of the free coefficients $c_J$ describing how the flow emanates from the fixed point. In general, this leads to relations among the low-energy couplings which could be tested by observations.

In the present work, we focus on a specific subsystem of \eqref{eq.ansmaster}, given by the action\footnote{The action \eqref{LagrangianfromASWST} falls into the class of Horndeski-type theories, characterized by the feature that the resulting equations of motion are of second order in the time-derivatives. This is a property of the approximation \eqref{LagrangianfromASWST} which does not extend to the full effective action $\Gamma_k$ though. Formulated differently, the condition that a theory is of Horndeski type is not preserved along the RG flow: starting from a Horndeski-type theory and integrating out quantum fluctuations will inevitably generate interactions which lead to higher-derivative terms in the equations of motion. 
}
\begin{equation}\label{LagrangianfromASWST}
    \Gamma_k = \int d^4x \, \sqrt{g} \left[\frac{2 \Lambda_k - R}{16 \pi \, G_k} + \frac{Z_k}{2} \, X + Z_k \, D_k \, R \, X\right].
\end{equation}
The dimensionless counterparts of the couplings appearing in this Lagrangian are
\be\label{dimlessdef}
g_k \equiv G_k \, k^2 \, , \quad \lambda_k = \Lambda_k \, k^{-2} \, , \qquad d_k \equiv D_k k^2 \, , \qquad \eta_s \equiv- k \p_k \ln(Z_k) \, . 
\ee
The fixed points identified for this subsystem are collected in Table \ref{Tab.1}. 
\begin{table}[t!]
	\centering
	\begin{tabular}{c|c||c|c|c||c|c|c||c}
	\hline
	 & $\eta_s$ & $g^*$ & $\lambda^*$ & $d^*$ & $\theta_3$ & $\theta_1$ & $\theta_2$ & \# \\
	\hline
	\small{GFP} & $0$ & $0$ & $0$ & $0$   & $-2$ & $2$ & $-2$ & 1 \\ 
 \hline	\hline
	\small{NGFP$_1$} & $\phantom{-}1.129$ & $0.668$ & $\phantom{-}0.211$ & $-0.737$ & $-3.198$ & \multicolumn{2}{c||}{$1.659 \pm 3.329i$} & 2 \\
	\hline	
    \small{NGFP$_2$} & $-0.459$ & $0.656$ & $\phantom{-}0.193$ & $\phantom{-}1.308$ & $-7.262$ & \multicolumn{2}{c||}{$1.743 \pm 2.938i$} & 2 \\
	\hline	
	\small{NGFP$_3$} & $-0.428$ & $0.657$ & $\phantom{-}0.209$ & $-0.19$ & $\phantom{-}2.324$ & \multicolumn{2}{c||}{$1.59 \pm 3.33i$} & 3 \\
	\hline
	\end{tabular}
\caption{\label{Tab.1} Fixed-point structure and critical exponents of the system \eqref{LagrangianfromASWST} obtained in \cite{Laporte:2021kyp}. The discussion is limited to the NGFPs where the RG flow admits a cross-over to the GFP. The column \# lists the number of free parameters available when the fixed point is used as a high-energy completion of the RG flow via the asymptotic safety mechanism.
}
\end{table}
The classical low-energy physics is obtained in the vicinity of the GFP.  

A phenomenological demand for choosing a particular NGFP for providing a phenomenologically viable UV-completion of the theory is that the RG flow emanating from the fixed point gives rise to RG trajectories which are complete in the sense that they extend from the quantum regime to a suitable classical phase at low energy. This selects the fixed points NGFP$_1$, NGFP$_2$, and NGFP$_3$. Two of these fixed points have negative $\theta_3$ values (NGFP$_1$ and NGFP$_2$). Thus, the asymptotic safety condition that RG trajectories should be captured by the fixed points in the regime where $\frac{k}{k_0} \gg 1$ demands $c_3=0$. The NGFP$_3$ has one more relevant direction. In this case  $c_3$ is an additional free parameter which should be fixed by observations.

\section{Scalar-tensor theories from RG improvement }
\label{sect.3}
At this point we are interested in the cosmologies supported by asymptotically safe models of the type \eqref{LagrangianfromASWST}. In order to investigate this question, we first transition from Euclidean to Lorentzian signature, so that 
\begin{equation}\label{LagrangianfromASWST2}
    \Gamma_k = \int d^4x \, \sqrt{-g} \left[\frac{R-2 \Lambda_k}{16 \pi \, G_k} - \frac{1}{2} \, X -  D_k \, R \, X\right].
\end{equation}
Here we have set $Z_k = 1$, since this is an inessential coupling. 
Moreover, we want to include the next layer of quantum corrections to this action. For this purpose, we resort to the method of RG improvements \cite{Bonanno:2006eu} which is frequently used in the context of asymptotically safe gravity phenomenology. 

In the latter context we point out, for the first time, that there are two conceptually different ways to perform the RG-improvement. The improvement can either be based on constructing ``leading quantum corrections'' at the NGFP or the GFP. So far, applications in terms of cosmology have been based on the former while the latter approach has not been employed in this context yet. Given that inflation is expected to occur at scales below the Planck scale, it is natural to expect that the corrections around the Gaussian fixed point may be to ones relevant to describe this process. Therefore, we contrast the two approaches in Sects.\ \ref{sect.31} and \ref{IRapproach}, respectively, before proceeding with the analysis based on the RG-improvement at the GFP in the latter sections.

\subsection{Expansions in local regions of the flow map}
\label{sect.30}
\begin{figure}[t!]
    \centering 
    \includegraphics[width=.77\linewidth]{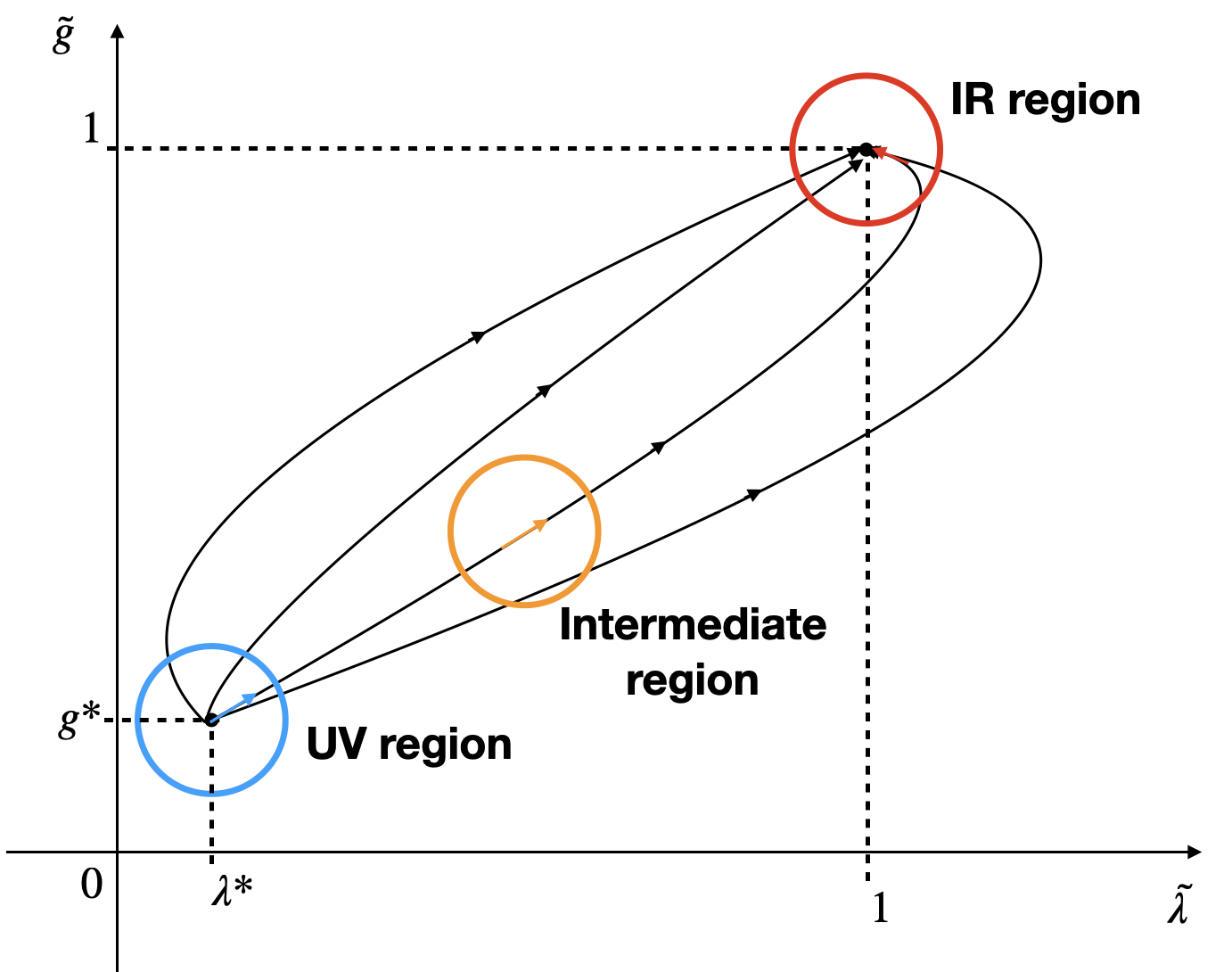}
    \caption{Schematic RG flow for the couplings $\tilde g_k$ and $\tilde \lambda_k$ interpolating between a NGFP in the UV  and a GFP in the IR. Expansions in the UV are indicated by the blue circle, expansions in the IR are indicated by the red circle. The typical energy scales associated with inflation are situated in the intermediate region (orange circle).}\label{fig:RG-flowRegions}
\end{figure}

The effective action resulting from a complete and asymptotically safe RG trajectory is supposed to be well-defined at all momentum scales. As an example of this generic situation, let us consider a two-dimensional projection of the RG flow captured by the scale-dependent couplings $\Lambda_k$ and $G_k$.
In the vicinity of a RG fixed point, their scale-dependence is given by
\be\label{FPscaling}
\begin{array}{lll}
\text{NGFP:} & G_k = g^* \, k^{-2} \, , \qquad & \Lambda_k = \lambda^* \, k^2 \, , \\
\text{GFP:} & G_k = G_0 \, , \qquad & \Lambda_k = \Lambda_0 \, . 
\end{array}
\ee
For the purpose of depicting the flow of these couplings in a compact way, 
let us define the dimensionless quantities
\bea
\tilde g_k=G_k \left( \frac{1}{G_0}+k^2\right), \qquad 
\tilde \lambda_k = \frac{\Lambda_k}{\Lambda_0+k^2}.
\eea
A RG trajectory connecting the NGFP as $k \rightarrow \infty$ to the GFP as $k\rightarrow 0$ then connects the points
\bea
\lim_{k\rightarrow \infty}\{\tilde g_k,\, \tilde \lambda_k\} = \{g^*,\; \lambda^*\} \, ,\qquad
\lim_{k\rightarrow 0}\{\tilde g_k,\, \tilde \lambda_k\} = \{1,\; 1\} \, .
\eea
A sketch of this situation is shown in Fig.\ \ref{fig:RG-flowRegions}.

\newpage
Let us now consider an ``experiment'' which is performed at energy scale $k_{obs}$. 
Ideally, the experiment can give information on the couplings at this scale by measuring $\tilde g(k_{obs})$ and $\tilde \lambda(k_{obs})$. 
If the experiment further covers some energy range
$k_{min}<k_{obs}<k_{max}$,  we could hope for additional information, namely the 
rate of change of the couplings in the vicinity of $k_{obs}$. 
This rate of change corresponds to the beta functions evaluated for the couplings measured at $k_{obs}$,
($ \beta_{\tilde\lambda}(\tilde g(k_{obs}),\tilde \lambda(k_{obs}))$ and $ \beta_{\tilde g}(\tilde g(k_{obs}),\tilde \lambda(k_{obs}))$), which arise from a local expansion of the flow at a fixed point along the RG trajectory.
In Fig.\ \ref{fig:RG-flowRegions}, this additional information is shown as the length and direction of the colored local arrows.
One can straightforwardly translate this dataset
 back into the more common notion of dimensionful couplings and beta functions
 $\left\{k_{obs}, \, G_{k_{obs}}, \,  \Lambda_{k_{obs}},\, \beta_{g}(g_{k_{obs}}, \lambda_{k_{obs}}), \,  \beta_{\lambda}(g_{k_{obs}}, \lambda_{k_{obs}})\right\}$.
Typical regions of interest for such an expansion are the deep UV ($k \rightarrow \infty$), the deep IR ($k \rightarrow 0$), or some
intermediate region. 
While there is a considerable theoretical interest in the blue UV region, it seems hard to imagine that one could obtain experimental information in this regime.
In contrast, the red IR regime is accessible from a phenomenological point of view \cite{Koch:2025yuz} and the orange intermediate region is situated in between these two extremes.

In the following subsection, we will explore
a higher-dimensional projection of the RG flow and discuss
the RG-improvement process in the UV region (blue) and the IR region (red).
\subsection{Expansion of the couplings around the NGFP}
\label{sect.31}
We start by rewriting the Lagrangian density \eqref{LagrangianfromASWST}  in terms of the dimensionless couplings \eqref{dimlessdef}
\begin{equation}\label{DimensionlessLagrangian}
    \mathcal{L}_k = \frac{k^2}{16 \pi \, g_k} \, \left(R-2 k^2 \, \lambda_k \right) - \frac{X}{2} - \frac{d_k}{k^2} \, RX.
\end{equation}
In this approximation, the beta functions give rise to one GFP and three NGFPs, see Table \ref{Tab.1}.
We can then use the linearized flow \eqref{gSolution} to obtain the $k$-dependence of the dimensionless couplings in the vicinity of a generic NGFP. Focusing on the phenomenologically interesting cases associated with fixed points NGFP$_1$, NGFP$_2$, and NGFP$_3$ where the stability coefficients contain a complex conjugate pair
\be
\theta_{1,2} = \theta^\prime \pm i \theta^{\prime\prime} \, , \quad \theta_3 \, , 
\ee
 the running of the dimensionless couplings at the linearized level is explicitly given by 
\begin{subequations}\label{couplingExpansion}
\begin{align}
\begin{split}\label{NewtonExpansion}
    g_k =& g^* + \left[\left(c_1 \cos{(\theta^{\prime\prime} t)} + c_2 \sin{(\theta^{\prime\prime} t)}\right) \, e^2_1 + \left(c_1 \sin{(\theta^{\prime\prime} t)} - c_2 \cos{(\theta^{\prime\prime} t)}\right) \, e^2_2\right] \left(\frac{k}{k_0}\right)^{-\theta^\prime}\\
    &+ c_3 \, e^2_{3} \left(\frac{k}{k_0}\right)^{-\theta_3},
\end{split}\\
\begin{split}\label{lambdaExpansion}
    \lambda_k =& \lambda^* + \left[\left(c_1 \cos{(\theta^{\prime\prime} t)} + c_2 \sin{(\theta^{\prime\prime} t)}\right) \, e^1_1 + \left(c_1 \sin{(\theta^{\prime\prime} t)} - c_2 \cos{(\theta^{\prime\prime} t)}\right) \, e^1_2\right] \left(\frac{k}{k_0}\right)^{-\theta^{\prime}}\\
    &+ c_3 \, e^1_{3} \left(\frac{k}{k_0}\right)^{-\theta_3},
\end{split}\\
\begin{split}\label{dExpansion}
    d_k =& d^* + \left[\left(c_1 \cos{(\theta^{\prime\prime} t)} + c_2 \sin{(\theta^{\prime\prime} t)}\right) \, e^3_1 + \left(c_1 \sin{(\theta^{\prime\prime} t)} - c_2 \cos{(\theta^{\prime\prime} t)}\right) \, e^3_2\right] \left(\frac{k}{k_0}\right)^{-\theta^{\prime}}\\
    &+ c_3 \, e^3_{3} \left(\frac{k}{k_0}\right)^{-\theta_3}.
\end{split}
\end{align}
\end{subequations}
Here the $c_J$, $J=1,2,3$ are the free coefficients describing how the underlying RG trajectory emanates from the fixed point. The case where Re$(\theta_3) < 0$ corresponds to a UV-irrelevant direction, the condition that the UV-completion should be provided by the fixed point indicates that the coefficient $c_3$ should be set to zero in order to ensure that one still describes RG trajectories which are captured by the NGFP as $k\rightarrow \infty$.

Eq.\ \eqref{DimensionlessLagrangian} contains the coupling $g_k$ in a denominator. Abbreviating 
\be 
\mathbb{A} \equiv \left[\left(c_1 \cos{(\theta^{\prime\prime} t)} + c_2 \sin{(\theta^{\prime\prime} t)}\right) \, e^2_1 + \left(c_1 \sin{(\theta^{\prime\prime} t)} - c_2 \cos{(\theta^{\prime\prime} t)}\right) \, e^2_2\right] \, , 
\ee
this inverse can formally be written as a geometric series in the linear perturbation around the fixed point
\begin{align}\label{Inversionformulag}
    \frac{1}{g_k} &= \frac{1}{g^*} \sum^{\infty}_{n=0} \frac{(-1)^n}{(g^*)^n}\left[\mathbb{A}\left(\frac{k}{k_0}\right)^{-\theta^{\prime}} + c_3 \, e^2_{3} \left(\frac{k}{k_0}\right)^{-\theta_3} \right]^n \, . 
\end{align}
Since Eq.\ \eqref{gSolution} is valid to first order in $c_J$, we consistently retain only terms linear in $c_J$ throughout. 
Thus
\begin{align}\label{Inversionformulagexp}
    \frac{1}{g_k}   &= \frac{1}{g^{*}} \left[1 - \frac{\mathbb{A}}{g_{*}} \left(\frac{k}{k_0}\right)^{-\theta^{\prime}} - \frac{c_3 \, e^2_{3}}{g_*} \left(\frac{k}{k_0}\right)^{-\theta_3} \right] + O(c_J^2) \, . 
\end{align}
Substituting Eqs.\ \eqref{lambdaExpansion}, \eqref{dExpansion}, and \eqref{Inversionformulagexp} into the Lagrangian \eqref{DimensionlessLagrangian} makes the scale-dependence of $\mathcal{L}$ in the vicinity of a NGFP explicit.

An RG-improvement identifies the coarse-graining scale $k$ with a physical quantity. A basic requirement is that the physical quantity appearing in this process is invariant with respect to a change of coordinates. In the context of cosmology this can be implemented in a covariant way by relating the IR cutoff $k$ with the Ricci scalar $R$ \cite{Weinberg:2009wa,Reuter:2005kb,Hindmarsh:2011hx,Bonanno:2015fga} 
\begin{equation}\label{IdentificationScale}
    k^2 = \xi \, R.
\end{equation}
Here, $\xi > 0$ is a dimensionless constant. The basic idea underlying such an identification is that the curvature $R$ provides an effective mass to quantum fluctuations. This effect suppresses the contributions of quantum fluctuations with momenta below this threshold. Thus the system may admit an effective description by the effective average $\Gamma_k[g]|_{k^2 = \xi R}$. The value of $\xi$ can be interpreted as balancing radiative and higher-order curvature corrections: when $\xi \ll 1$, the higher-order curvature corrections in the action (\ref{DimensionlessLagrangian}) become more relevant than the radiative corrections. On the contrary, if $\xi \gg 1$, the radiative corrections are important while the curvature corrections become negligible \cite{Copeland:2013vva,Weinberg:2009wa}. The insertion of Eqs.\ \eqref{IdentificationScale} together with \eqref{lambdaExpansion}, \eqref{dExpansion} and \eqref{Inversionformulag} into the action \eqref{DimensionlessLagrangian} gives 
\begin{align}\label{UVRGImprovementLagrangian}
    \mathcal{L}^{UV}_{RG} =& \, s_0 \, R^2 + s_1 \, R^{2-\frac{\theta^{\prime}}{2}} \, \cos{\left(\frac{\theta^{\prime\prime}}{2} \, \ln{\left(\frac{R}{k^2_0}\right)}\right)} + s_2 \, R^{2-\frac{\theta^{\prime}}{2}} \, \sin{\left(\frac{\theta^{\prime\prime}}{2} \, \ln{\left(\frac{R}{k^2_0}\right)}\right)} \nonumber\\
    & + s_3 \, R^{2-\frac{\theta_3+\theta^{\prime}}{2}} \, \cos{\left(\frac{\theta^{\prime\prime}}{2} \, \ln{\left(\frac{R}{k^2_0}\right)}\right)} + s_4 \, R^{2-\frac{\theta_3+\theta^{\prime}}{2}} \, \sin{\left(\frac{\theta^{\prime\prime}}{2} \, \ln{\left(\frac{R}{k^2_0}\right)}\right)}\nonumber\\
    & + s_5 \, R^{2-\theta_3} + s_6 \, R^{2-\frac{\theta_3}{2}} + s_7 \, R^{2-\theta^{\prime}} + s_8 \, R^{2-\theta^{\prime}} \, \cos{\left(\theta^{\prime\prime} \, \ln{\left(\frac{R}{k^2_0}\right)}\right)}\nonumber\\
    & + s_9 \, R^{2-\theta^{\prime}} \, \sin\left({\theta^{\prime\prime} \, \ln{\left(\frac{R}{k^2_0}\right)}}\right) + O(X)\, .
\end{align}
Here we have shifted the reference scale $k_0^2 \mapsto k_0^2/\xi$. Moreover, contributions containing the additional scalar field are not written explicitly and denoted collectively by the expression $O(X)$. The coefficients $s_i$ are listed in Table \ref{Tab.2}. 
\begin{table}[t!]
\centering
\renewcommand{\arraystretch}{1.6}
\begin{tabular}{ll}
coefficient & value \\ \hline \hline
        $s_0$ & $\frac{\xi \, (1-2 \, \xi \, \lambda^*)}{16 \pi \, g^*}$ \\
        $s_1$ & $\frac{\xi \, k^{\theta^{\prime}}_0}{16 \pi \, (g^*)^2} \, \left[(2 \xi \, \lambda^* - 1) \, (c_1 \, e^2_1 - c_2 \, e^2_2) - 2 \xi \, g^* \, (c_1 \, e^1_1 - c_2 \, e^1_2)\right]$ \\
        $s_2$ & $\frac{\xi \, k^{\theta^{\prime}}_0}{16 \pi \, (g^*)^2} \, \left[(2 \xi \, \lambda^* - 1) \, (c_2 \, e^2_1 - c_1 \, e^2_2) - 2 \xi \, g^* \, (c_2 \, e^1_1 + c_1 \, e^1_2)\right]$ \\
         $s_3$ & $\frac{\xi^{2} \, k^{\theta_3+\theta^{'}}_0}{8 \pi \, (g^*)^2} \, \left[c_1 \, c_3 \, (e^1_3 \, e^2_1 + e^1_1 \, e^2_3) - c_2 \, c_3 \, (e^1_3 \, e^2_2 + e^1_2 \, e^2_3)\right]$ \\
        $s_4$ & $\frac{\xi^{2} \, k^{\theta_3+\theta^{'}}_0}{8 \pi \, (g^*)^2} \, \left[c_1 \, c_3 \, (e^1_3 \, e^2_2 + e^1_2 \, e^2_3) + c_2 \, c_3 \, (e^1_3 \, e^2_1 + e^1_1 \, e^2_3)\right]$ \\
        $s_5$ & $\frac{c^2_3 \, e^1_3 \, e^2_3 \, \xi^{2} \, k^{2 \theta_3}_0}{8 \pi \, (g^*)^2}$ \\
        $s_6$ & $\frac{\xi \, k^{\theta_3}_0}{16 \pi \, (g^*)^2} \, c_3 \left[-e^2_3 + 2 \xi \left(\lambda^* \, e^2_3 - g^* \, e^1_3\right)\right]$ \\
        $s_7$ & $\frac{\xi^{2} \, k^{2 \theta^{\prime}}_0}{16 \pi \, (g^*)^2} \, \left(c^2_1 + c^2_2\right) \left(e^1_1 \, e^2_1 + e^1_2 \, e^2_2\right)$ \\
        $s_8$ & $\frac{\xi^{2} \, k^{2 \theta^{\prime}}_0}{16 \pi \, (g^*)^2} \, \left[(c^2_1-c^2_2) (e^1_1 \, e^2_1 + e^1_2 \, e^2_2) - 2 c_1 \, c_2 (e^1_1 \, e^2_2 + e^1_2 \, e^2_1)\right]$ \\
        $s_9$ & $\frac{\xi^{2} \, k^{2 \theta^{\prime}}_0}{16 \pi \, (g^*)^2} \, \left[(c^2_1-c^2_2) (e^1_1 \, e^2_2 + e^1_2 \, e^2_1) + 2 c_1 \, c_2 (e^1_1 \, e^2_1 - e^1_2 \, e^2_2)\right]$ \\ \hline \hline
 $$
\end{tabular}
\caption{\label{Tab.2} Expansion coefficients appearing in the RG-improved Lagrangian based on a NGFP. For a fixed point where Re($\theta_3) < 0$, the relation simplify since asymptotic safety dictates $c_3=0$. In this case, $s_3 = s_4=s_5=s_6=0$ and one recovers the case analyzed in \cite{Bonanno:2018gck} up to a different alignment of the (now three-dimensional) eigenvectors associated with the relevant directions.}
\end{table}

The RG-improvement turns the scale-dependent action \eqref{DimensionlessLagrangian} into an $f(R)$-type Lagrangian, whose interactions and couplings depend on the position of the fixed point $\{g^*,\lambda^*\}$, its stability coefficients $\theta^\prime, \theta^{\prime\prime}, \theta_3$, and the RG trajectory chosen when flowing away from the fixed point $\{c_1, c_2, c_3\}$. In \cite{Bonanno:2018gck} these features have been used to place potential bounds on these quantities based on the condition that the system gives rise to an inflationary phase compatible with observations. The resulting bounds on the stability coefficients were then translated into conditions on number of matter fields needed for the construction to give results compatible with observations.

\subsection{Expansion of the couplings in the infrared region}\label{IRapproach}
The cross-over between the scaling regions dictated by the NGFP and GFP occurs at the Planck scale. Inflation is expected to happen at energy scales below this scale. This motivates a different strategy for carrying out the RG-improvement by ``incorporating the leading quantum correction'' to the low-energy Lagrangian associated with the GFP. In this way, the RG-improvement is based on a "bottom-up" instead of the usual "top-down" approach in the sense of  energy scales. In this case, the procedure for determining the $k$-dependence of the couplings given in Eq.\ \eqref{couplingExpansion} is adjusted as follows. 

The three dimensionful couplings in the action \eqref{LagrangianfromASWST2} can be expanded analytically in a Taylor series for small values $k$ (i.e., the low-energy region)
\be\label{IRexpansion}
        G_k = \sum^{\infty}_{n=0} G_n \, k^n 
\, , \qquad 
        \Lambda_k = \sum^{\infty}_{n=0} \Lambda_n \, k^n
  \, , \qquad 
        D_k = \sum^{\infty}_{n=0} D_n \, k^n \, . 
  \ee
Here $G_0,\Lambda_0,D_0$ are free coefficients giving the value of the couplings at $k=0$. The couplings $G_k, \Lambda_k$, and $D_k$ have an even mass dimension. Since in $d=4$, the beta functions \cite{Laporte:2021kyp} are analytic in the couplings, all expansion coefficients multiplying an odd power of $k$ are zero. By substituting the ansatz \eqref{IRexpansion} into the beta functions and solving the resulting hierachy order by order in powers of $k^2$, the coefficients $G_n, \Lambda_n, D_n$ for $n$ even are fully determined in terms of $G_0, \Lambda_0, D_0$. The explicit form of the expansion up to terms of order $k^8$ reads
\begin{subequations}\label{IRCouplingExpansion}
\begin{align}
    \begin{split}\label{IRCouplingExpansiona}
        G_k =& \, G_0 - \frac{13 \, G^2_0}{12 \pi} \, k^2 + \frac{169 G^3_0 \, \Lambda_0 - 30 \pi \, G^2_0 - 24 \pi \, D_0 \, \Lambda_0 \, G^2_0}{144 \pi^2 \, \Lambda_0} \, k^4 \\
        &- \frac{G^2_0 \left[336 \pi^2 + 2197 \, G^2_0 \, \Lambda^2_0 - 2 \pi \, G_0 \, \Lambda_0 \left(197 + 312 \, D_0 \, \Lambda_0\right)\right]}{1728 \pi^3 \, \Lambda^2_0} \, k^6 + O(k^8) \, ,
    \end{split}\\
    \begin{split}\label{IRCouplingExpansionb}
        \Lambda_k =& \, \Lambda_0 - \frac{13 \, G_0 \, \Lambda_0}{12 \pi} \, k^2 + \frac{169 \, G^2_0 \, \Lambda_0 - 93 \pi \, G_0 - 24 \pi \, D_0 \, \Lambda_0 \, G_0}{144 \pi^2} \, k^4 \\
        &- \frac{G_0 \left[696 \pi^2 + 2197 \, G^2_0 \, \Lambda^2_0 - \pi \, G_0 \, \Lambda_0 \left(1213 + 624 \, D_0 \, \Lambda_0\right)\right]}{1728 \, \pi^3 \, \Lambda_0} \, k^6 + O(k^8) \, ,
    \end{split}\\
    \begin{split}\label{IRCouplingExpansionc}
        D_k =& \, D_0 - \frac{G_0}{48 \pi \, \Lambda_0} \, k^2 - \frac{G_0 \, D_0}{24 \pi \, \Lambda_0} \, k^4 \\
        &+ \frac{G_0 \left(252 \pi-1512 \pi \, D_0 \, \Lambda_0 + 53 G_0 \, \Lambda_0 - 432 \pi \, {D_0}^2 \, {\Lambda_0}^2\right)}{20736 \pi^2 \, \Lambda^3_0} \, k^6 + O(k^8) \, .
    \end{split}
\end{align}
\end{subequations}
Note that the coefficient $D_0$ enters the expansion at order $k^4$ only. The RG-improved action is then obtained by substituting the IR expansion \Eqref{IRCouplingExpansion} into \eqref{LagrangianfromASWST2} and eliminating $k$ through the cutoff-identification \eqref{IdentificationScale}. Retaining all terms containing up to six spacetime-derivatives this results in the IR-improved Lagrangian
\begin{align}\label{IRRGimprovedLagrangian}
    \mathcal{L}^{IR}_{RG} = \frac{1}{2 \kappa^2} \, \left( R -2 \, \Lambda_{0} + \frac{1}{6 \, m^2} \, R^2 + \frac{\lambda}{3 \, m^4} \, R^3\right) - X \left(\frac{1}{2} + D_0 \, R - \alpha \, R^2 \right) .
\end{align}
Here we adjusted the notation to the one typically employed when analyzing inflation in $f(R)$-type theories. The relation of the parameters to the free coefficients $\{\Lambda_0, G_0, D_0, \xi\}$ is given by
\be\label{CoupStaLik}
\kappa^2 = 8 \pi G_0 \, , \quad
        m^2 = \frac{4 \pi}{\xi \, G_0 \, \left(26 + 21 \, \xi\right)} \, , \quad 
        \lambda = \frac{2 \pi \left(5 + 10 \xi + 4 \, D_0 \, \Lambda_0\right)}{\left(26 + 21 \, \xi\right)^2 \, G_0 \, \Lambda_0} \, , \quad
        \alpha = \frac{\xi \, G_0}{48 \pi \, \Lambda_0} \, . 
\ee

Structurally, the RG-improvement induces higher-order curvature terms $R^2, R^3, \cdots$ with integer-valued powers. The four-derivative part in the gravitational sector of \eqref{IRRGimprovedLagrangian} is actually independent of $D_0$. The non-minimal matter coupling enters at order $R^3$ only. Thus, if we seek to bound this coupling through inflationary dynamics, it is necessary to retain the expansion to this order in the curvature.

It is worth noting that the parameter $D_0$ entering the IR-improved Lagrangian is not a freely adjustable coupling in the full asymptotically safe theory. Rather, it is determined by the coefficients $c_J$ specifying how the RG trajectory emanates from the UV fixed point, and is therefore subject to the predictive constraints of the asymptotic safety mechanism. From this perspective, a specific numerical value of $D_0$ does not constitute a fine-tuning in the traditional sense, but rather corresponds to the selection of a particular RG trajectory in the UV-complete theory. The shift symmetry of the scalar kinetic term, proven to be preserved under the RG flow in \cite{Laporte:2021kyp}, further ensures that the value of $D_0$ is technically natural: the symmetry forbids the generation of operators that would radiatively destabilize this coupling, so that a large value of $D_0$ is stable against quantum corrections below the Planck scale.

Having established these results, it is interesting to compare the UV- and IR-based improvements \eqref{UVRGImprovementLagrangian} and \eqref{IRRGimprovedLagrangian}. A notable difference is that the improvement at a NGFP induces non-analytic terms and powers of the Ricci scalar which are set by the stability coefficients of the fixed point. Hence, this type of improvement knows about these characteristic features of the UV-completion. This property is absent in the IR-improvement. In this case one obtains integer powers in the scalar curvature. Moreover, the construction is agnostic about the UV-completion. The interactions generated by the improvement process depend on the free low-energy couplings only. The conditions provided by the UV-completion are not (fully) taken into account. Specifically, the enhanced predictive power associated with NGFP$_1$ and NGFP$_2$ leads to one further relation between these couplings once this information is included. This feature will then manifest itself in one additional relation which allows to express $D_0$ in terms of the two other free parameters $\Lambda_0$ and $G_0$.

At this point a critical remark about the scope and limitations of the RG-improvement procedure are in order. It is well known, that the Coleman-Weinberg potential encoding one-loop corrections to a scalar $\phi^4$-theory can be obtained through this procedure \cite{Coleman:1973jx,Migdal:1973si,Gross:1973ju,Pagels:1978dd,Matinyan:1976mp}. On the other hand, it has been shown that an RG-improvement based on Newton's coupling can not be used to obtain graviton-mediated scattering amplitudes obtained within effective field theory \cite{Donoghue:2019clr}. The latter does not come as a surprise, as the amplitudes depend on \emph{two gravitational form factors} (each being an independent function of the squared momentum of the process) while an RG-improvement based on a scale-dependent Newton's coupling gives one function of a single variable only \cite{Knorr:2019atm,Bonanno:2020bil,Draper:2020knh}. Despite these quantitative limitations, there is a key feature of asymptotic safety that is correctly captured by the RG-improvement process leading to \eqref{IRRGimprovedLagrangian}: the couplings that appear in the low-energy effective field theory description are not independent \cite{Saueressig:2023irs}. The asymptotic safety mechanism entails that they are determined in terms of the free coefficients $c_J$ describing how a RG trajectory emanates from the UV-fixed point. This feature is inherited by the RG-improved Lagrangian where the couplings of the higher-derivative terms can be read as functions of the free parameters via the relation \eqref{CoupStaLik}. Conceptually, it is then an interesting question whether there are observations that could test such relations. From this perspective, the RG-improvement is an interesting laboratory which allows to address such a question before embarking on the (technically highly involved) task of determining these relations from first-principle computation.

\section{Representations of Scalar-Tensor theories of $F(R,\phi)$-type}
\label{sect.4}
The RG-improved actions \eqref{UVRGImprovementLagrangian} and \eqref{IRRGimprovedLagrangian} fall into the class of generalized $F(R)$-theories of the form
\begin{equation}\label{GenericAction}
    \bar{S} = \frac{1}{2 \, \kappa^2} \int d^4x \, \sqrt{-g} \left[F(R,\phi) - \kappa^2 g^{\mu\nu} (\partial_{\mu}\phi) (\partial_{\nu}\phi)\right] \, . 
\end{equation}
Here, $F(R,\phi)$ is a function on the scalar curvature $R$ and the scalar field $\phi$. Their characteristic property is that they possess a global shift symmetry $\phi \rightarrow \phi + c$, where $c$ is a real constant. Moreover, the IR-improved action contains higher-order scalar curvature terms. This implies that, besides the spacetime metric and the scalar $\phi$, they possess one additional scalar degree of freedom, the scalaron $\rho$.

In order to study the dynamics resulting from these models, it is convenient to convert \eqref{GenericAction} to the Einstein frame where the scalaron is minimally coupled to gravity. This conversion is performed in two steps. In the first step, we recast the $F(R,\phi)$-theory into a form that makes the contribution of the scalaron manifest. In the second step, we apply a conformal transformation of the metric so that the scalaron becomes minimally coupled. The construction follows \cite{Canko:2019mud}.

\subsection{Equivalence between scalar tensor and $F(R,\phi)$ gravity}
We start by recasting Eq.\ \eqref{GenericAction} into the form
\begin{equation}\label{AlternativeAction}
    S = \frac{1}{2 \, \kappa^2} \int d^4x \, \sqrt{-g} \left[F(\bar{\Phi},\phi) - \kappa^2 X + \psi \left(R -\bar{\Phi}\right)\right] \, . 
\end{equation}
Here $\bar{\Phi}$ is a scalar field and $\psi$ acts as a Lagrange multiplier. The equation of motion for $\psi$ is 
\begin{equation}\label{eom1}
    \frac{\delta S}{\delta \psi} = 0 \quad \Rightarrow \quad \bar{\Phi} = R \, . 
\end{equation}
Substituting Eq.\ \eqref{eom1} into \eqref{AlternativeAction} shows the equivalence between $\bar{S}$ and $S$. The variation of $S$ with respect to $\bar{\Phi}$ gives
\begin{equation}\label{Inversion}
    \frac{\delta \, S}{\delta \, \bar{\Phi}} = 0 \quad \Rightarrow \quad \frac{\partial F(\bar{\Phi},\phi)}{\partial \bar{\Phi}} = \psi \, .  
\end{equation}
Assuming that this relation can be solved for $\bar{\Phi}$ allows to express $\bar{\Phi}$ as a function of $\psi, \phi$
\be\label{inv2}
\bar{\Phi} = \bar{\Phi} (\psi,\phi) \, . 
\ee

At this point, we return to the IR-improved Lagrangian \eqref{IRRGimprovedLagrangian}. Comparing this Lagrangian to the generic action \eqref{AlternativeAction} shows that in this case 
\begin{equation}\label{Fbarphi}
    F(\bar{\Phi},\phi) = -2 \Lambda_0 + \left(d_1 \, \bar{\Phi} + d_2 \, \bar{\Phi}^2 + d_3 \, \bar{\Phi}^3\right) + (d_4 + d_5 \, \bar{\Phi}) \, \bar{\Phi} \, X. 
\end{equation}
The $d_i$, $i=1,...,5$ can be read off from \eqref{IRRGimprovedLagrangian} and are determined in terms of the free parameters of the construction
\be\label{diidentification}
d_1 = 1 \, , \; \; d_2 = \frac{1}{2m^2} \, , \; \; d_3 = \frac{\lambda}{3m^4} \, , \; \; d_4 = \textcolor{blue}{-}2 \kappa^2 D_0 \, , \; \; d_5 = 2 \kappa^2 \alpha \, .  
\ee
The polynomial structure of $F(\bar{\Phi},\phi)$ allows to determine the function \eqref{inv2} in closed form 
\begin{equation}\label{PhiSolution}
    \bar{\Phi}_{\pm} = - \frac{(d_2 + d_5 \, X) \pm \sqrt{(d_2 + d_5 \, X)^2 - 3 d_3 \, (d_1+d_4 \, X -  \psi)}}{3 \, d_3} \, .
\end{equation}
The sign ambiguity can be resolved by insisting on a well-defined limit $d_3 \rightarrow 0$.  Physically, this limit implements the idea that the $R^3$-term should be treated as a perturbation of the dynamics governed by the $R^2$-interaction. This selects the $\bar{\Phi}_-$-branch.

When this term is inserted into the action \Eqref{AlternativeAction}, the Lagrangian becomes a second-order polynomial in $\bar{\Phi} \equiv \bar{\Phi}_-$ 
\be\label{EqnoExpansion}
\begin{split}
    S = \frac{1}{2 \kappa^2} \int d^4 x \sqrt{-g} \,  \Big[& \psi R -2 \, \Lambda_0 - \kappa^2 X \\ 
    & +\frac{1}{3} \left(2 d_1 + d_2 \bar{\Phi}-2 \psi\right) \bar{\Phi} +\frac{2}{3} (d_4 + \frac{1}{2} d_5 \bar{\Phi}) \bar{\Phi}  X \Big] \, .
\end{split}
\ee
This action  corresponds to the scalar-tensor representation of the RG-improved Lagrangian \eqref{IRRGimprovedLagrangian}, highlighting that there are two scalar degrees of freedom $\phi$, $\psi$. While $\phi$ is minimally coupled to gravity and does not appear in the potential, $\psi$ is non-minimally coupled and has no kinetic term.

Notably, the transformation to the scalar-tensor formulation induces an infinite tower of momentum-dependent interaction terms for the scalar fields. Slow-roll inflation is caused by a scalar field ``rolling slowly'' in a suitable scalar potential. A typical assumption made in this regime is that the kinetic terms associated with the scalars are negligible. This suggests an expansion of \eqref{EqnoExpansion} in powers of $X$. For the second line in \eqref{EqnoExpansion}, this gives
\be
\begin{split}
& \frac{1}{3}  \left(2 d_1 + d_2 \bar{\Phi}-2 \psi\right) \bar{\Phi} +\frac{2}{3} (d_4 + 2 d_5 \bar{\Phi}) \bar{\Phi}  X 
 \\ 
& \qquad  = \frac{(\mathbb{D} - d_2)(6 d_1 d_3 -d_2^2  - 6 d_3 \psi +  d_2 \, \mathbb{D})}{27 d_3^2} 
\\
& \qquad \quad + \frac{
 (\mathbb{D} - d_2) (3 d_3 d_4 - 2 d_2 d_5 + 2 d_5 \, \mathbb{D})}{
 9 \, d_3^2} \, X \, + O(X^2) \, , 
\end{split}
\ee
where $\mathbb{D} \equiv \left( d_2^2 - 3 d_1 d_3 + 3 d_3 \psi \right)^{1/2}$. The expansion coefficients then contribute a potential term for $\psi$ as well as a field-dependent prefactor of the $\phi$-kinetic term
\be
\begin{split}
 \mathbb{P}(\psi) =& 2\Lambda_0 - \frac{(\mathbb{D} - d_2)(6 d_1 d_3 -d_2^2  - 6 d_3 \psi +  d_2 \, \mathbb{D})}{27 d_3^2}   \, , \\
\mathbb{K}(\psi) =& \kappa^2 - \frac{
 (\mathbb{D} - d_2) (3 d_3 d_4 - 2 d_2 d_5 + 2 d_5 \, \mathbb{D})}{
 9 \, d_3^2} \, . 
\end{split}
\ee
With these definitions, Eq.\ \eqref{EqnoExpansion} takes the form
\begin{equation}\label{ActionJordanFrame}
    S = \frac{1}{2 \kappa^2} \int d^4x \sqrt{-g} \left(\psi R   - \mathbb{K}(\psi) X - \mathbb{P}(\psi) + O(X^2) \right) \, . 
\end{equation}
This concludes recasting the IR-improved Lagrangian into scalar-tensor form.

\subsection{Conformal transformation of the metric field}

The non-minimal coupling of $\psi$ can be removed by a conformal transformation of the metric field. This transformation from the Jordan frame to the Einstein frame gives rise to a kinetic term for the scalaron.\footnote{While the mathematical equivalence between these formulations is guaranteed, physical concepts may have a different appearance in the two frames.  
Points as the conservation of energy conditions in both frames of certain modified theories of gravity or the preservation of the weak equivalence principle in the Jordan frame but not in the Einstein frame \cite{Faraoni:2006fx} have been topics of debate for a long time. In particular, in Brans-Dicke's theory, the scalar field violates all of the energy conditions in the Jordan frame but satisfies them in the Einstein frame. Additionally, massive particles in the Einstein frame do not follow timelike geodesics due to an extra force proportional to the gradient of the scalar field. Hence, the Weak Equivalence Principle is satisfied in
the Jordan frame but not in the Einstein frame due to the coupling of the scalar field to ordinary matter \cite{Faraoni:2010pgm}. Moreover, the equivalence of the theories in the quantum regime is still under debate \cite{Kamenshchik:2014waa}. In this work, we stipulate that the observables do not depend on the frame one computes them; therefore, the physical equivalence of the notion of conformal frames in cosmology holds at least in the classical regime \cite{Kubota:2011re, Chiba:2013mha,Jarv:2014hma}.}
The metrics in the Jordan frame ($g_{\mu\nu}$) and Einstein frame ($\tilde{g}_{\mu\nu}$) are related through the conformal factor $\Omega$ 
\begin{equation}
    g_{\mu\nu} = \Omega^{-2} \, \tilde{g}_{\mu\nu} \, , \qquad  g^{\mu\nu} = \Omega^2 \, \tilde{g}^{\mu\nu} \, , \qquad \Omega^2 \equiv \psi \, .
\end{equation}
The transformation of the determinant of the metric, and the relation between the Ricci scalars in both frames are given by \cite{Wald:1984rg} 
\be\label{RicciJRicciE}
\begin{split}
     \sqrt{-g} = \Omega^{-4} \sqrt{-\tilde{g}} \, , \qquad
    R = \Omega^2 \tilde{R} - \frac{3}{2} \frac{\tilde{g}^{\mu\nu}}{\Omega^2} (\partial_\mu \Omega^2) (\partial_\nu \Omega^2) + 3 \tilde{g}^{\mu\nu} \partial_\mu \partial_\nu \Omega^2.
\end{split}
\ee
The last term in the $R$-transformation yields a surface term and will be disregarded in the sequel. Applying this transformation to Eq.\ \eqref{ActionJordanFrame} then yields the action in the Einstein frame 
\begin{align}\label{ActionEinsteinFrame}
    \tilde{S} = \frac{1}{2 \kappa^2} \int d^4x \sqrt{-\tilde{g}} \left[\tilde{R}-\frac{3}{2} \frac{\tilde{g}^{\mu\nu}}{\psi^2} (\partial_\mu \psi) (\partial_\nu \psi) -  \psi^{-1} \, \mathbb{K}(\psi) \, \tilde{X} - \psi^{-2} \, \mathbb{P}(\psi)\right] \, ,
\end{align}
where $\tilde{X} \equiv \tilde{g}^{\mu\nu} X_{\mu\nu}$. Finally, the kinetic term of the scalar field $\psi$ can be cast into canonical form by redefining
\begin{equation}\label{FieldRedefinition}
    \psi = \exp{\left[\sqrt{\frac{2}{3}} \, \kappa \rho\right]} \, . 
\end{equation}
Introducing
\begin{equation}\label{M&V}
    \mathbb{M}(\rho) = \frac{1}{\kappa^2} \exp{\left(\sqrt{\frac{2}{3}} \kappa \, \rho\right)} \mathbb{K}(\rho) \, , \qquad  \mathbb{V}(\rho) = \frac{1}{2\kappa^2} \, \exp{\left(-2 \, \sqrt{\frac{2}{3}} \kappa \, \rho\right) \, \mathbb{P}}(\rho) \, ,
\end{equation}
brings the action into its canonical form
\begin{equation}\label{ActEinsCan}
    \tilde{S} = \int d^4x \sqrt{-\tilde{g}} \left[\frac{\tilde{R}}{2 \kappa^2}-\frac{1}{2} \tilde{g}^{\mu\nu} (\partial_\mu \rho) (\partial_\nu \rho) - \frac{1}{2} \, \mathbb{M}(\rho) \, \tilde{X} - \mathbb{V}(\rho)\right] \, .
\end{equation}

At this stage, one readily verifies that the potential \eqref{M&V} agrees with the scalar potential underlying Starobinsky inflation if the $R^3$-contribution is switched off. Evaluating $\mathbb{V}(\rho)$ in the limit $d_3 \rightarrow 0$ and setting $\Lambda_0 =0$ yields
\be\label{StarobinskylikePotential}
    \left. \mathbb{V}(\rho) \right|_{d_3 = \Lambda_0 = 0} =  \frac{m^2}{4\kappa^2} \, \left(1 - \exp{\left(- \sqrt{\frac{2}{3}} \, \kappa \rho\right)}\right)^2 \, .
\ee
This is the potential obtained in \cite{Starobinsky:1980te}.

\begin{figure}[t!]
    \centering
    \includegraphics[width=.7\linewidth]{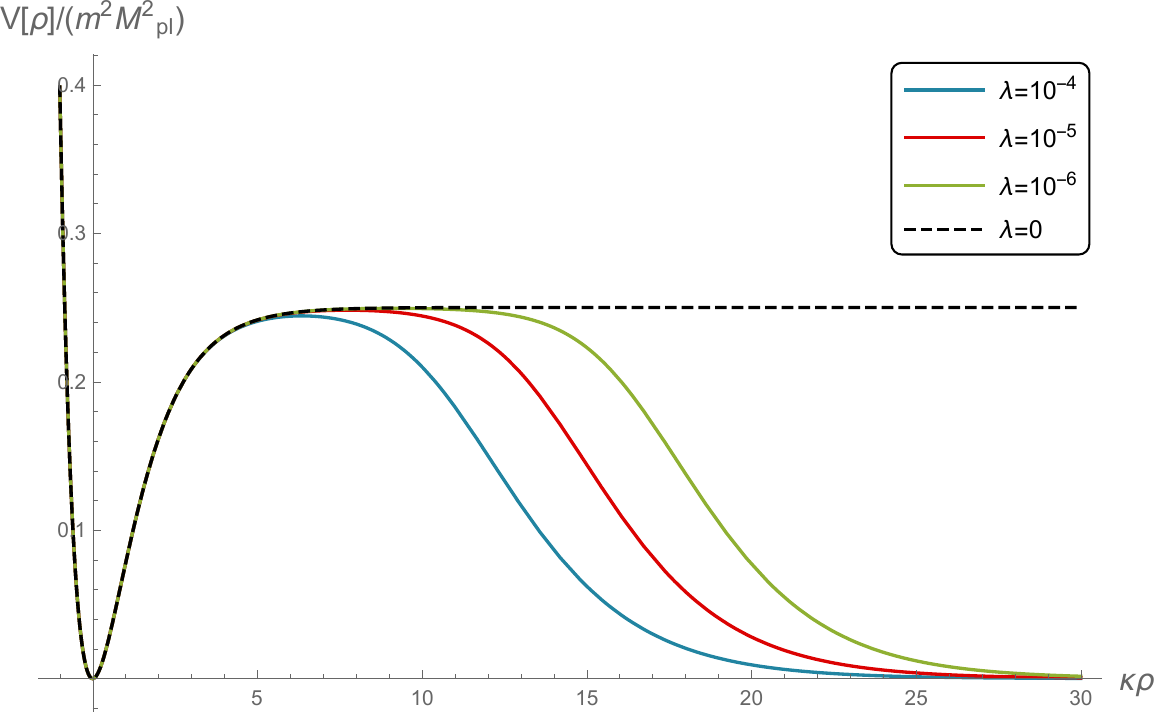}
    \caption{Potential $\mathbb{V}(\rho)$ as a function of the scalar field $\kappa \, \rho$ for different values of $\lambda$. The smaller the magnitude of $\lambda$, the larger the coordinate of the maximum of the plateau. A viable slow-roll evolution of the scalaron field requires that the maximum of the plateau is sufficiently flat.}
    \label{fig:Potential}
\end{figure}

The behavior of the potential $\mathbb{V}(\rho)$ as a function of the couplings $\lambda$, $m^2$, and $\Lambda_0$ is shown in Fig.\ \ref{fig:Potential}. Here, we added \eqref{StarobinskylikePotential} as a dashed black reference line. 
The figure illustrates that the key effect of including the $R^3$ term is to shrink the extension of the plateau region of $\mathbb{V}(\rho)$ to a finite $\rho$-interval. For $\rho \rightarrow \infty$, the potential vanishes in the presence of the $R^3$-term. Ultimately, it is this shrinking mechanism which allows to bound the parameter $D_0$ from the size of the cubic correction term.

\section{Covariant multi-field formalism applied to the RG improved Lagrangian}
\label{sect.5}

The action \Eqref{ActEinsCan} can be seen as a special case of a generalized non-linear sigma model of multi-field inflation \cite{Mori:2017caa}
\begin{equation}\label{MultifieldAction}
    S = \int d^4x \, \sqrt{-\tilde{g}} \left[\frac{R}{2 \, \kappa^2} - \frac{1}{2} \zeta_{IJ} \, \tilde{g}^{\mu\nu} (\partial_\mu \phi^I) \, (\partial_\nu \phi^J) - \mathbb{V}(\phi^I)\right] \, .
\end{equation}
In the above action, the field-space indices $I$ and $J$ label the two scalar fields, $\phi^I = (\phi,\rho)$, and  
$\zeta_{IJ}$ represents the metric on the curved field space manifold characterizing the geometry of the target space spanned by
the fields. Given the model presented in \Eqref{ActionEinsteinFrame}, the metric and its inverse are given by
\begin{equation}\label{NLSMmetric}
    \zeta_{IJ} = 
    \begin{pmatrix}
    1 & 0\\
    0 & \mathbb{M}(\rho)
    \end{pmatrix} \hspace{.3 cm},\hspace{.3 cm} \zeta^{IJ} = 
    \begin{pmatrix}
    1 & 0\\
    0 & \mathbb{M}(\rho)^{-1} 
    \end{pmatrix}.
\end{equation}
Before proceeding, it is worth commenting on the role of the multi-field formalism in the present context. The action \eqref{ActEinsCan} contains two scalar degrees of freedom, the scalaron $\rho$ and the original scalar field $\phi$. However, since $\phi$ enters neither the potential $V(\rho)$ nor the field-space metric \eqref{NLSMmetric} directly (it only appears through its kinetic term weighted by $\mathbb{M}(\rho)$) the inflationary trajectory lies entirely along the $\rho$-direction, with $\phi$ constituting a flat direction. As a consequence, the multi-field dynamics reduces effectively to a single-field problem for the purposes of computing the slow-roll parameters and the CMB observables. The multi-field formalism is nevertheless retained here for two reasons. First, it provides a manifestly covariant treatment of the two-field system that remains valid beyond the slow-roll approximation and facilitates the analysis of perturbations, including potential isocurvature modes sourced by fluctuations along the $\phi$-direction. Second, once suitable fermionic content and Yukawa couplings are included in the asymptotic safety framework, a non-trivial scalar potential for $\phi$ is expected to be generated, at which point the full multi-field structure of the formalism will become essential. The present setup should therefore be understood as establishing the framework for such future extensions.

In this field covariant formulation, it is natural to define a covariant differential operator $\mathcal{D}_t$ by the action on vectors in the field space $X^I$ 
\begin{equation}
    \mathcal{D}_t \, X^I = \Dot{X}^I + \Gamma^I_{AB}\, \Dot{\phi}^A \, X^B \, ,
\end{equation}
where we have denoted $\dot{(\,)} \equiv \frac{d}{dt}(\,)$. The metric $\zeta_{IJ}$ defines the Christoffel connection in curved field space 
\begin{equation}
    \Gamma^K_{IJ} = \frac{1}{2} \zeta^{KL} \left(\partial_I \, \zeta_{LJ} + \partial_J \, \zeta_{IL} - \partial_L \, \zeta_{IJ}\right) \, .
\end{equation}
\Eqref{NLSMmetric} defines the Christoffel symbols in terms of $\mathbb{M}(\rho)$
\begin{align}\label{CSs}
    \Gamma^\rho_{\rho\rho} = 0 \hspace{.3 cm},\hspace{.3 cm} \Gamma^\rho_{\rho\phi} = \Gamma^\rho_{\phi\rho} = 0 \hspace{.3 cm},\hspace{.3 cm} \Gamma^{\rho}_{\phi\phi} = - \frac{1}{2} \, \frac{\partial\mathbb{M}(\rho)}{\partial \rho}\nonumber\\
    \Gamma^{\phi}_{\rho\rho} = 0 \hspace{.3 cm},\hspace{.3 cm} \Gamma^\phi_{\rho\phi} = \Gamma^\phi_{\phi\rho} = \frac{1}{2} \frac{\partial \log \mathbb{M}(\rho)}{\partial \rho} \hspace{.3 cm},\hspace{.3 cm} \Gamma^\phi_{\phi\phi} = 0 \, .
\end{align}
Under the assumption of a homogeneous and isotropic spacetime described by the Friedmann-Lemaitre-Robertson-Walker (FLRW) metric
\begin{equation}\label{FRWmetric}
    ds^2 = -dt^2 + a(t)^2 \, d\Omega^2_3 \, ,
\end{equation}
with $a(t)$ being the scale factor and $d\Omega^2_3$ a spatially flat three-dimensional space, the variation of the action \eqref{MultifieldAction} with respect to the metric $\tilde{g}_{\mu\nu}$ leads to the Friedmann equation 
\begin{equation}\label{FriedmannEqMultifieldFormalism}
    H^2 = \frac{\kappa^2}{6} \left(\zeta_{IJ} \, \dot{\phi}^I \, \dot{\phi}^J + 2 \mathbb{V}(\phi^I)\right) \, ,
\end{equation}
where $H=\frac{\dot{a}}{a}$ is the Hubble rate. Here we have assumed that the scalar fields share the symmetries of the spacetime. The continuity equation reads
\begin{equation}\label{ContinuityEqMultifieldFormalism}
    \dot{H} = -\frac{\kappa^2}{2} \, \zeta_{IJ} \, \dot{\phi}^I \, \dot{\phi}^J \, .
\end{equation}
The equations of motion for the homogeneous fields $\phi^I(t)$ are
\begin{equation}\label{SFequations}
    \mathcal{D}_{t} \dot{\phi}^I + 3 H \dot{\phi}^I + \zeta^{IJ} \, \frac{\partial \, \mathbb{V}(\phi^I)}{\partial \, \phi^{J}} = 0 \, .
\end{equation}
Within the covariant multi-field formalism in the FLRW background, deviations from the de Sitter space are measured by the slow-roll parameters\footnote{In \cite{Mori:2017caa}, the second slow-roll parameter $\eta$ has been defined as $\eta = 2 \left(\epsilon + \frac{\zeta_{IJ} \, \dot{\phi}^I \, \mathcal{D}_t \, \dot{\phi}^J}{H \, \zeta_{KL} \, \dot{\phi}^K \, \dot{\phi}^L}\right)$. Here, we use a different definition since $\eta$ carries an extra factor of 2 when a unique field spans the metric of the curved field space compared with the canonical single-field case. While this does not represent a problem when the respective changes are implemented in the spectral tilt, we will stick to the canonical definition of the CMB's observables.} \cite{He:2018gyf,Ghilencea:2018rqg}
\begin{subequations}\label{SlowRollPar}
\begin{equation}\label{slowroll1}
    \epsilon = \frac{\kappa^2}{2} \, \frac{\zeta_{IJ} \, \dot{\phi}^I \, \dot{\phi}^J}{H^2} \, ,
\end{equation}
\begin{equation}\label{slowroll2}
    \eta_{\|} = \frac{\zeta_{IJ} \, \dot{\phi}^I \, \mathcal{D}_t \, \dot{\phi}^J}{H \, \zeta_{KL} \, \dot{\phi}^K \, \dot{\phi}^L} \, .
\end{equation}
\end{subequations}
Slow-roll inflation requires that $\epsilon\ll 1$ , $\eta_{\|} \ll 1$. 

\section{Properties of the potential and slow-roll inflation}
\label{sect.6}

The analysis of the potential in Fig.\ \ref{fig:Potential} shows three distinct regions: a flat potential in the asymptotic limit $\rho \gg 1$, the plateau in the central region, and the end of scalaron inflation and the corresponding reheating era close to the origin. Inflation occurs when the inflaton rolls to the left of the maximum position of the plateau (see Fig.\ \ref{fig:Potential}), leaving the asymptotic region unphysical. The main goal of this section is to analyze the region to the left with respect to the maximum of the plateau where the slow-roll inflation takes place with the RG improved action obtained in Sect.\ \ref{IRapproach}. Observational constraints obtained by the combined Planck, WMAP, and BICEP/Keck CMB observations allow setting boundaries on the values of $\xi$ and $D_0$ and then selecting preferred RG trajectories that give suitable scenarios for inflation. 

\subsection{Equations of motion}

Motivated by the shape of the potential in Fig.\ \ref{fig:Potential}, it turns out useful to introduce the dimensionless variable $\sigma$ 
\begin{equation}\label{sigmaDefinition}
    \sigma \equiv \exp{\left[-\sqrt{\frac{2}{3}} \, \kappa \, \rho\right]} \, ,
\end{equation}
because it is small  during the slow-roll inflation. The potential will be studied under the assumptions of a homogeneous and isotropic Universe described by the FRLW metric \Eqref{FRWmetric}. The Christoffel symbols in \Eqref{CSs} define the Friedmann equation \eqref{FriedmannEqMultifieldFormalism} of our model
\begin{align}
    H^2 = \frac{\kappa^2}{6} \left(\dot{\rho}^2 + \mathbb{M}(\rho) \, \dot{\phi}^2 + 2 \mathbb{V}(\rho)\right) \, .
\end{align}
While the continuity equation only depends on the kinetic terms for the scalar fields $\phi(t), \rho(t)$
\begin{align}
    \dot{H} &= -\frac{\kappa^2}{2} \, \left(\dot{\rho}^2 + \mathbb{M}(\rho) \dot{\phi}^2\right) \, .
\end{align}
The equations of motion for $\phi$ and $\rho$ are given by \footnote{Note that \Eqref{SFequations} admits an additional term proportional to the derivative of the potential with respect to $\phi$. Since $\phi$ constitutes a flat direction in the potential, this term vanishes.} 
\begin{subequations}
\begin{equation}
    \Ddot{\rho} + 3 H \, \dot{\rho} - \frac{\dot{\phi}^2}{2} \, \frac{\partial \mathbb{M}(\rho)}{\partial \rho} + \frac{\partial \mathbb{V}(\rho)}{\partial \rho}= 0,
    \end{equation}
\begin{equation}
    \Ddot{\phi} + 3 H \, \dot{\phi} + \frac{\dot{\phi} \, \dot{\rho}}{2} \, \frac{\partial \, \log \mathbb{M}(\rho)}{\partial \, \rho} =0 \, .
\end{equation}
\end{subequations} 

%
\subsection{Slow-roll inflation}
%
Now we will study the physics occurring in the plateau, where physical inflation and the slow-roll paradigm can be applied. In this region, the application of the first slow-roll approximation $\{\dot{\phi}^2,\dot{\rho}^2\} \ll \mathbb{V}$ leads to the simplified Friedmann equation  
\begin{equation}\label{Friedmann1Simplified}
    H^2 \simeq \frac{\kappa^2}{3} \, \mathbb{V} \, .
\end{equation}
Considering the second slow-roll approximation $\Ddot{\rho} \ll H \, \dot{\rho}$, the $\rho$-equation of motion reads (the Klein-Gordon equation of $\phi$ is not necessary since the potential does not depend on this field) 
\begin{equation}\label{Friedmann2Simplified}
    3 H \, \dot{\rho} \simeq -\frac{\partial \mathbb{V}}{\partial \rho} \, . 
\end{equation}
The first step consists in analyzing the properties of the potential. For $\mathbb{V}$ to be well-defined for every real number $\lambda$, the asymptotic limits
\begin{itemize}
\centering
    \item $\rho \rightarrow \infty \hspace{.3 cm} \Rightarrow \hspace{.3 cm} \lambda \rightarrow 0$
    \item $\rho \rightarrow -\infty \hspace{.3 cm} \Rightarrow \hspace{.3 cm} \lambda < \frac{1}{4}$ \, ,
\end{itemize}
impose the condition $0 \leq \lambda < 0.25$. With a positive $\lambda$, all terms involved in the potential are positive; therefore, the absence of ghosts and negative values of the potential are ensured. Within these bounds, the maximum of the potential occurs at 
\begin{equation}\label{MaximumOfThePlateau}
    \rho_c = \frac{1}{\kappa} \, \sqrt{\frac{3}{2}} \, \log \left(4 + \sqrt{\frac{3}{\lambda}}\right) \, .
\end{equation}
To arrive at \Eqref{MaximumOfThePlateau}, we have ignored the effects produced by the cosmological constant term since its effect in the values of the potential is highly suppressed by its value at the infrared, $\Lambda_0 \simeq 10^{-84} \, \text{GeV}^2$. From the structure of $\rho_c$, one infers that the position of the maximum of the plateau depends entirely on the value of $\lambda$. When the effects of $D_0$ become important, the position of $\rho_c$ approaches values close to zero. Insofar as $\lambda$ takes small values, the plateau becomes wider and wider until at the limit $\lambda \rightarrow 0$ one recovers the shape of the potential observed in Starobinsky inflation, as shown in \Eqref{StarobinskylikePotential}. The main goal of this section is to obtain the slow-roll parameters in the regime ($\rho < \rho_c$). In this sector, $\sigma \ll 1$ and $\lambda$ can be expressed as 
\begin{equation}\label{lambdaitosc}
    \lambda = \frac{3}{e^{2 \psi_c} - 8 e^{\psi_c} + 16} \simeq \frac{3}{e^{2 \psi_c}} \equiv 3 \sigma^2_c \, ,
\end{equation}
where $\psi_c = \psi(\rho=\rho_c)$. Under the assumption $\sigma \sim \sigma_c \ll 1$, $\mathbb{V}$ can be written in terms of \Eqref{M&V} and approximate by 
\begin{align}\label{PotentialApprox}
    \mathbb{V}(\sigma) &= \frac{\sigma \, m^2}{48 \kappa^2 \, \lambda^2} \left[1 - \sqrt{1 - 4 \lambda \, \sigma^{-1} \, (\sigma - 1)}\right] \left[-\sigma + 8 \lambda \, (\sigma-1)+\sigma \, \sqrt{1-4 \lambda \, \sigma^{-1} \, (\sigma-1)}\right]\nonumber\\
    &\simeq \frac{m^2}{4 \kappa^2} \, \left(1 - 2 \sigma - 2  \, \frac{\sigma^2_c}{\sigma}\right) + \mathcal{O}\left(\sigma^2\right) \, .
\end{align}
Additionally 
\begin{equation}\label{FirstDerPotentialApprox}
    \mathbb{V}^{'}(\sigma) \equiv \frac{d \, \mathbb{V}}{d \, \sigma} \frac{d \, \sigma}{d \, \rho} \simeq \frac{m^2}{2 \kappa} \, \sqrt{\frac{2}{3}} \, \left(\sigma - \frac{\sigma^2_c}{\sigma}\right) + \mathcal{O}\left(\sigma^2\right) \, ,
\end{equation}
\begin{equation}\label{SecondDerPotentialApprox}
    \mathbb{V}^{''}(\sigma) \equiv \frac{d \, \mathbb{V}^{'}}{d \, \sigma} \frac{d \, \sigma}{d \, \rho} \simeq -\frac{m^2}{3} \,  \left(\sigma + \frac{\sigma^2_c}{\sigma}\right) + \mathcal{O}\left(\sigma^2\right) \, .
\end{equation}
The slow-roll parameters defined in \Eqref{SlowRollPar} can be expressed as a function of the potential and its derivatives using the equations of motion for the scalar fields $\rho$ and $\phi$. 
Therefore, an expansion up to second-order in $\sigma$ retains the relevant information of $\epsilon$ and $\eta_{\|}$ in the slow-roll regime. 
The general multi-field formalism introduced in Sect.\ \ref{sect.5}, $\epsilon$ and $\eta_{\|}$ depend on the multi-field potential. The flat $\phi$-direction reduces the 
inflationary trajectory to the single-field case 
$\epsilon = \frac{1}{2 \kappa^2} \left(
\frac{\mathbb{V}'}{\mathbb{V}}
\right)^2$, 
$\eta_{\|} = \frac{1}{2 \kappa^2} \left(\frac{\mathbb{V} \, \mathbb{V}^{''}-\mathbb{V'}^2}{\mathbb{V}^2}\right)$. The first parameter $\epsilon$ in \Eqref{slowroll1} as a function of $\sigma$ reads
\begin{align}\label{slowroll1sigma}
    \epsilon &\simeq \frac{4}{3} \, \left(\sigma - \frac{\lambda}{3 \, \sigma}\right)^2 \, \left(1-2 \sigma - \frac{2 \lambda}{3 \sigma}\right)^{-2} + \mathcal{O}(\sigma^3)\nonumber\\
    &\simeq \frac{4}{3} \, \sigma^2 \, \left(1 - \frac{\sigma^2_c}{\sigma^2}\right)^2 + \mathcal{O}\left(\sigma^3\right) \, .
\end{align}
Note that the first non-vanishing contribution in \Eqref{slowroll1sigma} appears at second order in $\sigma$. For the second slow-roll parameter $\eta_{\|}$ \Eqref{slowroll2} 
\begin{align}\label{slowroll2sigma}
    \eta_{\|} &\simeq -\frac{4}{3} \, \left(1-2 \sigma - \frac{2 \lambda}{3 \sigma}\right)^{-2} \, \left[\left(\sigma + \frac{\lambda}{3 \sigma}\right) \, \left(1 - 2 \sigma - \frac{2 \lambda}{3 \sigma}\right) - \left(\sigma - \frac{\lambda}{3 \sigma}\right)^2\right] + \mathcal{O}(\sigma^3)\nonumber\\
    &\simeq - \frac{4}{3} \, \left(\sigma + \frac{\sigma^2_c}{\sigma}\right) - \frac{8}{3} \, \left(\sigma + \frac{\sigma^2_c}{\sigma}\right)^2 +  \mathcal{O}(\sigma^3) \, .
\end{align}
Contrary to what we observed in $\epsilon$, $\eta_{\|}$ starts at $\mathcal{O}(\sigma)$. 
The last task consists in expressing the pair $\{\epsilon,\eta_{\|}\}$ as a function of the number of e-folds instead of the $\sigma$-variable to make contact with the observations in the next chapter. The number of e-folds is defined as 
\begin{equation}
    N = \int^{t_{end}}_t H(\bar{t}) \, d\bar{t} \, .
\end{equation}
To perform the integral, it is convenient to trade the time $t$ for $\sigma$, employing the chain rule 
\begin{equation}\label{dt}
    dt = \frac{d t}{d \sigma} d \sigma = - \, \sqrt{\frac{3}{2}} \, \frac{d \sigma}{\kappa \, \sigma \, \dot{\rho}} \, .
\end{equation}
The value of $\dot{\rho}$ can be extracted from \Eqref{Friedmann1Simplified} and \Eqref{Friedmann2Simplified}
\begin{align}\label{DotRho}
    \dot{\rho} = -\frac{1}{3 \, \kappa} \, \sqrt{\frac{3}{\mathbb{V}}} \, \frac{\partial \mathbb{V}}{\partial \rho} \, .
\end{align}
By inserting \Eqref{PotentialApprox} and \Eqref{FirstDerPotentialApprox} into \Eqref{DotRho} and plugging the resulting expression back in \Eqref{dt} 
\begin{align}
    d t = \frac{3}{2 \sigma \, m} \, \sqrt{3 \left(1 - 2 \, \sigma - 2 \, \frac{\sigma^2_c}{\sigma}\right)} \, \left(\sigma - \frac{\sigma^2_c}{\sigma}\right)^{-1} \, d \sigma \, .
\end{align}
Finally, the Friedmann equation \Eqref{Friedmann1Simplified} allows to write the Hubble constant in the definition of $N$ in terms of $\sigma$ and $\sigma_c$
\begin{equation}\label{NumberOfefolds}
    N = \frac{3}{4} \, \int^{\sigma_{end}}_{\sigma} \frac{1 - 2 \bar{\sigma} - 2 \sigma^2_c \, \bar{\sigma}^{-1}}{\bar{\sigma} \, \left(\bar{\sigma}-\sigma^2_c \, \bar{\sigma}^{-1}\right)} \, d \bar{\sigma} \, .
\end{equation}
Here, $\sigma_{end}$ corresponds to the variable \Eqref{sigmaDefinition} evaluated at the end of inflation, defined by $\epsilon = 1$. Integrating \Eqref{NumberOfefolds} leads to a non-invertible expression due to the emergence of a hyperbolic arctan function that cannot be inverted within the approximations employed. To get $\sigma(N)$, we notice that $\sigma$ at the end of inflation takes values of the order $\mathcal{O}(10^{-1}-1)$ for the values of $\lambda$ displayed in Fig. \ref{fig:Potential}, while $\sigma_c \ll 1$. This means that the quantity $\tau_{end} \equiv \frac{\sigma_c}{\sigma_{end}} \ll 1$ can be employed in \Eqref{NumberOfefolds}, making the evaluation of the upper  integration limit more tractable. Furthermore, the quantity $\tau \equiv \frac{\sigma_c}{\sigma}$ is small in the slow-roll regime. By expressing $N$ in terms of $\tau$, the integration becomes 
\begin{align}\label{FullN}
    N =& \frac{3}{4} \, \int^\tau_{\tau_{end}} \, \frac{\bar{\tau} - 2 \, \sigma_c - 2 \, \bar{\tau}^2 \, \sigma_c}{\sigma_c \, \bar{\tau} \, (1 - \bar{\tau}^2)} \, d \bar{\tau}\nonumber\\
    =& \frac{3}{8 \sigma_c} \, \Big\{ \log\left(\frac{1+\tau}{1-\tau}\right) - \log\left(\frac{1+\tau_{end}}{1-\tau_{end}}\right) + 4 \sigma_c \log\left[(1+\tau) \, (1-\tau)\right]\nonumber\\
    &-4 \sigma_c \, \log\left[(1-\tau_{end}) \, (1+\tau_{end})\right] - 4 \sigma_c \, \log\left(\frac{\tau_{end}}{\tau}\right) \Big\} \, .
\end{align}
Since $\tau_{end} \ll 1$, the second and fourth terms vanish. Under the slow-roll approximation, $\sigma_c \ll 1$ implies that the third term is also 0. For the fifth term, the logarithm goes to infinity, while the term multiplying it goes to zero. In the region where the slow-roll inflation takes place, $\sigma_c$ is always positive, as shown below \Eqref{Friedmann2Simplified}, so the limit $\lim_{x \rightarrow 0^{+}} x \, \log(x) = 0$ proves that the fifth term in \Eqref{FullN} vanishes. The only term surviving in $N$ is the first one
\begin{align}
    N \simeq \frac{3}{8 \, \sigma_c} \, \log\left(\frac{1+\tau}{1-\tau}\right) \, .
\end{align}
Inverting the previous relation gives $\xi(N)$
\begin{equation}
    \tau = \frac{\exp\left(\frac{8}{3} \, \sigma_c \, N\right)-1}{\exp\left(\frac{8}{3} \, \sigma_c \, N\right)+1} \hspace{.3 cm}\Rightarrow\hspace{.3 cm} \sigma = \sigma_c \, \frac{\exp\left(\frac{8}{3} \, \sigma_c \, N\right)+1}{\exp\left(\frac{8}{3} \, \sigma_c \, N\right)-1} \, ,
\end{equation}
and correspondingly, the slow-roll parameters \Eqref{slowroll1sigma} and \Eqref{slowroll2sigma} in terms of $N$ read
\begin{subequations}\label{slowrollN}
\begin{equation}\label{epsilonN}
    \epsilon \simeq \frac{64 \, \sigma^2_c}{3} \, \frac{\exp{\left(\frac{16}{3} \, \sigma_c \, N\right)}}{\left(\exp{\left(\frac{16}{3} \, \sigma_c \, N\right)}-1\right)^2} \, ,
\end{equation}
\begin{equation}\label{etaN}
    \eta_{\|} \simeq -\frac{8 \, \sigma_c}{3} \, \frac{ 4 \sigma_c - 1 + (1+4 \sigma_c) \, \exp{\left(\frac{32}{3} \, \sigma_c \, N\right)}}{\left(1-\exp{\left(\frac{16}{3} \, \sigma_c \, N\right)}\right)^2} \, .
\end{equation}
\end{subequations}
These are the expressions for the slow-roll parameters necessary for the analysis of the next section. Note that the leading order terms in an expansion in the region $\lambda \rightarrow 0$ (or, equivalently, $N \sigma_c \rightarrow 0$) reproduce the slow-roll parameters associated with Starobinsky inflation.

\section{Contrast with observations and bounds on $\xi$, $D_0$}
\label{sect.7}
\begin{figure}[!h]
    \centering
    \includegraphics[width=0.8\linewidth]{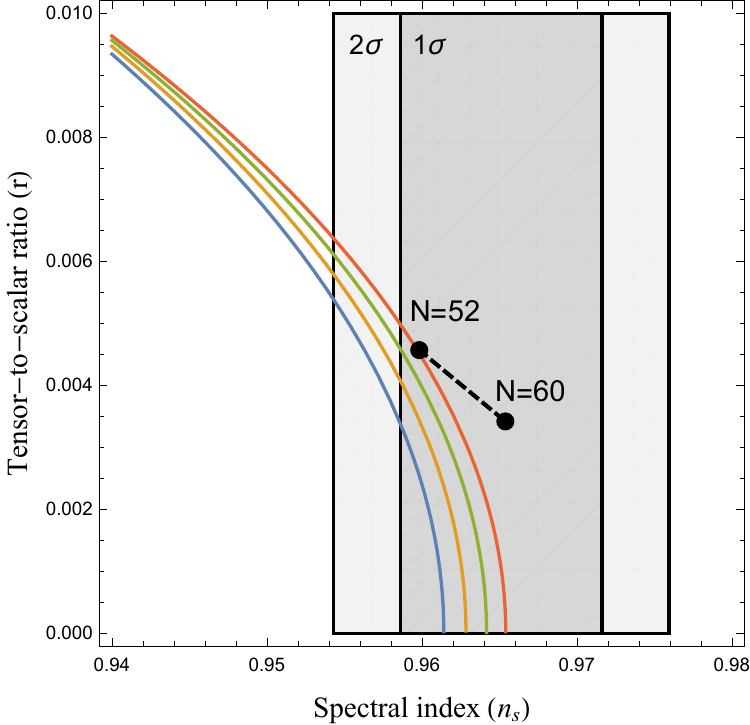}
    \caption{Predictions for the spectral index ($n_s$) and tensor-to-scalar ratio ($r$) in the two-field model described by Eq.\ \eqref{ActEinsCan} for $N \in [52,60]$. The blue, orange, green, and red lines correspond to $N=52$, $N=55$, $N=58$, and $N=60$, respectively. For reference purposes, we have included the Starobinsky case with black dots for the same range of e-folds. The grey regions in the figure denote the $68\%$ and $95\%$ confidence levels for the combined Planck satellite 2018 TT,TE,EE+lowE+lensing with a pivot scale $k=0.002 \, \text{Mpc}^{-1}$. Interestingly, a significant portion of each curve falls within the favored region of the observational data, even when we lower the maximum limit of $r$ to 0.01 for better visualization.
    }
    \label{fig:ObsCon}
\end{figure}

We assume that $G_0$ and $\Lambda_0$ are fixed by post-inflationary cosmology and astrophysics. The scalar-tensor theory  (\ref{IRRGimprovedLagrangian}) therefore has two remaining free parameters, the dimensionless quantity $\xi$ arising from the identification scale and the dimensionful infrared coupling $D_0$ associated with the $RX$ interaction. This freedom can be restricted by looking at the current constraints on inflationary theory from observations of CMB fluctuations. The power spectra of the scalar and tensorial perturbations describe the distribution of the temperature fluctuations in the CMB. This spectrum is characterized by two quantities: the spectral index $n_s$ that governs the scale dependence of the power spectrum against scalar perturbations, and the tensor to scalar ratio $r$, representing the suppression of the primordial spectrum of tensor perturbations against the scalar ones. In the slow-roll approximation, these quantities are
\begin{subequations}\label{r&ns}
\begin{align}
    \begin{split}
        n_s &= 1 + 2 \, \eta_{\|} - 4 \, \epsilon,
    \end{split}\\
    \begin{split}
        r &= 16 \, \epsilon,
    \end{split}
\end{align}
\end{subequations}
where $\eta_{\|}$ and $\epsilon$ are given by \Eqref{slowrollN}. Fig.\ \ref{fig:ObsCon} presents the model-generated curves in the $(n_s,r)$ plane that satisfy the observational constraints, as determined by the equation \Eqref{ActEinsCan}, for a given value of $\lambda$. These curves were generated by setting the coordinate of the plateau maximum as a function of the $N$ and $n_s$ and varying the number of e-folds in the range $52\leq N \leq 60$. The theoretical predictions made in Sect.\ \ref{sect.6} for the range of values of the spectral index are shown to be in excellent agreement with the observational constraints imposed by the Planck satellite in both plots. We find that for all values of $\sigma_c$ of order $O(10^{-4})$ or less, the predicted values of $n_s$ and $r$ fall within the marginalized joint 95{\%} confidence levels obtained by the Planck 2018 mission for the full range $52 \leq N \leq 60$.

The observational constraints on the CMB observables set stringent limits on the values of $D_0$ in the context of the $R^3$-inflation model. To extract the allowed values of $D_0$, we refer to Table \ref{Tab.D0values} for a range of $\sigma_c$. In order to reproduce the correct amplitude of scalar perturbations $\mathcal{A}_s \sim 2.1 \times 10^{-9}$ for $R^2$-inflation, we fix the value of the parameter $\xi$ to be of order $\mathcal{O}(10^{-5})$. This choice leaves the infrared value of the coupling associated with the non-minimal interaction of the Ricci scalar with the scalar kinetic term as the only free parameter in the model.  The requirement of a sufficiently extended inflationary plateau, as characterized by $\sigma_c \leq 10^{-4}$, then implies via \Eqref{lambdaitosc} and the relation \Eqref{CoupStaLik} that $D_0 \, \Lambda_0 \simeq -1.25$, with corrections that become negligible in this regime as shown in Table \ref{Tab.D0values}. Since $\Lambda_0 \sim 10^{-84} \, \text{GeV}^2$ is fixed by post-inflationary cosmology, this translates into a concrete bound on the magnitude of $D_0$. The tensor-to-scalar ratio is correspondingly predicted to lie in the range $2.1 \times 10^{-3} \lesssim r \lesssim 3.5 \times 10^{-3}$ for $52 \leq N \leq 60$, well within current Planck bounds $r < 0.056$ \cite{Planck:2018jri} and accessible to next-generation experiments such as LiteBIRD \cite{Kamionkowski:2015yta,2019JLTP..194..443H}.

The consistency conditions imposed by the inflationary dynamics require a sizeable negative value of $D_0$, specifically 
\begin{equation}\label{eq.final-result}
D_0 \, \Lambda_0 \simeq -1.25 \, . 
\end{equation}
This holds across the range of $\sigma_c$ explored in Table \ref{Tab.D0values}. At first sight, this appears to represent a significant fine-tuning of the infrared coupling associated with the non-minimal $RX$ interaction. However, two observations might mitigate this concern. First, within the asymptotic safety framework $D_0$ is not a free parameter but is determined by the RG trajectory connecting the UV fixed point to the infrared, as discussed in Sect.\ \ref{IRapproach}. The requirement \eqref{eq.final-result} then translates into a selection criterion on the admissible RG trajectories rather than a tuning of an independent low-energy parameter. Second, the shift symmetry of the scalar kinetic term preserved under the RG flow \cite{Laporte:2021kyp} ensures that this value is technically natural in the sense of 't Hooft: no radiative corrections are generated that would drive $D_0$ away from its required value.

The quantitative determination of which RG trajectories satisfy this condition, and whether they are generic or exceptional within the fixed-point structure of Table 1, is an important open question. At the level of the present IR-improvement, this cannot be determined from first principles; a definitive assessment requires a full computation within the asymptotic safety program along the lines of \cite{Saueressig:2024ojx,Silva:2024wit}. Crucially however, it is precisely the connection established in this work between the non-minimal gravity-matter coupling $D_0$ and inflationary observables that makes such a future computation meaningful: our results identify which values of $D_0$ are phenomenologically viable, providing a concrete target for first-principle asymptotic safety calculations.
\begin{table}[t!]
	\centering
	\begin{tabular}{|c|c|c|c|c|c|}
	\hline
	$\sigma_c$ & $N_{min}$ & $N_{max}$ & $r_{min} \, \left(\times 10^{-3}\right)$ & $r_{max} \, \left(\times 10^{-3}\right)$ & $D_0 \, \Lambda_0$ \\
        \hline
	$0.005$ & $60.6307$ & $97.1712$ & $1.97$ & $0.56037$ & -1.25\\
	\hline
        $0.003$ & $53.3747$ & $71.8759$ & $2.974 $ & $1.5622 $ & -1.25\\
	\hline
	$0.001$ & $50.7584$ & $65.7112$ & $3.471 $ & $2.0631 $ & -1.25\\
	\hline
	$10^{-4}$ & $50.4661$ & $65.0742$ & $3.5303 $ & $2.1218 $ & -1.25\\
	\hline
        $10^{-5}$ & $50.4632$ & $65.0679$ & $3.5341 $ & $2.1256 $ & -1.25\\
        \hline
        $10^{-6}$ & $50.4631$ & $65.0679$ & $3.5342 $ & $2.1267 $ & -1.25\\
        \hline
	\end{tabular}
\caption{\label{Tab.D0values} Minimum and maximum values of the number of e-folds and the tensor-to-scalar ratio for various maximum values of the potential $\sigma_c$ obtained while adhering to the constraints on the spectral index $n_s$ derived from the combined Planck 2018 lensing data. The convergence of the results for $\sigma_c \lesssim 10^{-4}$ confirms that the inflationary predictions of the model are effectively independent of the precise value of the plateau parameter in this regime.
}
\end{table}
%

\section{Conclusions}
\label{sect.conclusions}
In this work, we studied specific models for scalar field inflation within the framework of scalar-tensor theories. Motivated by recent results obtained within the gravitational asymptotic safety program \cite{Laporte:2021kyp}, we imposed a global shift symmetry on the scalar field. Starting from a Horndeski-type theory, we then generated higher-derivative interactions through an RG improvement procedure. These interactions are expected to appear naturally from first principle computations, and the RG improvement should be read as a shortcut on understanding the underlying phenomenological implications without first doing the quantum computation.

In contrast to \cite{Bonanno:2018gck}, our RG improvement procedure is based on the Gaussian fixed point. The resulting action then has a similar structure as the one encountered in effective field theory. In particular only interactions with an integer power of the curvature tensors are created in the process. The key feature of the construction is that the couplings of the higher-derivative gravitational terms are given as functions of the free parameters appearing in the initial action. In particular, the gravity-matter coupling $D_0$, attributed to the non-minimal coupling of the spacetime curvature and the scalar kinetic term, also moves into the higher-derivatives terms in the gravitational sector. Specifically, the $R^3$-coupling contains this initial parameter. In this way, it is possible to actually constrain the non-minimal gravity-matter coupling from the inflationary dynamics. We stress that, while the constraints generated in this work may not be the ones arising from a first principle computation, \emph{the asymptotic safety program is expected to give rise to precisely such relations} \cite{Codello:2007bd,Benedetti:2009rx}. In connection to the fixed point structure reported in Table \ref{Tab.1} the constraints on $D_0$ imply that one could falsify a UV-completion by NGFP$_1$ and NGFP$_2$ where the value of $D_0$ is predicted from asymptotic safety. Conversely, it can be used to pinpoint phenomenologically viable RG trajectories emanating from NGFP$_3$ where $D_0$ is a free parameter.

The physics mechanism producing the constraints on $D_0$ is rather generic. In our construction, inflation is driven by the scalaron potential mainly generated by the $R^2$ contribution. The higher-order curvature terms including $R^3$ must then be subleading in the sense that the region of the scalar potential where inflation takes place is not significantly altered, c.f.\ Fig.\ \ref{fig:Potential}. Based on present-day observational data this leads to the constraints reported in Sect.\ \ref{sect.7}. In other words it is the extremum in the potential that sets the slow-roll parameters. The subleading terms associated with $D_0$ affect this extremum. It is this feature that bounds the value of $D_0$ ultimately.

The next generation of CMB experiments will test the validity of the phenomenology range of $R^2$ inflation and its higher derivative corrections. In this sense, note that in Fig. \ref{fig:ObsCon}, we have lowered the $r$-limit to $0.01$. This is considerably lower than the current upper limit $r<0.056$ provided by the Planck mission \cite{Planck:2018jri}, thus, all curves inside the grey regions are in good agreement with the current data. However, future satellite missions advocated the detection of the B-mode pattern \cite{Seljak:1996gy,Kamionkowski:1996zd}, the special type of polarization induced by an anisotropic stretching of
the spacetime due to the presence of a gravitational wave background, will tighten the present cosmological bounds.  The
(JAXA) LiteBIRD mission will, for instance, constrain $r <
0.002$ at $95\%$ confidence level, or otherwise determine $r$ up
to an error smaller than $0.001$ \cite{Kamionkowski:2015yta,2019JLTP..194..443H}. These bounds will impose even stronger constraints on the value on $D_0$ of our model. To impose infrared bounds over all the infrared couplings mediating the interactions of the scalar-tensor model studied in \cite{Laporte:2021kyp}, one has to go beyond the slow-roll inflation to reheating. The advantage of looking at the end of inflation lies in further contact with future ground-based experiments like POLARBEAR \cite{POLARBEAR:2019kzz}, CMB-S4 \cite{CMB-S4:2016ple}, CLASS \cite{2014SPIE.9153E..1IE}, Simons Observatory \cite{SimonsObservatory:2018koc},  the next generation BICEP array \cite{BICEP3:2016pqy} and the South Pole Telescope \cite{SPT-3G:2014dbx} to restrict even further the parameter space that involves additional free quantities. Since the reheating period involves nonperturbative, nonequilibrium effects and depends on the interaction processes between the inflaton and matter fields, this complex scenario will be investigated in future work. 

A key feature of our analysis is that the additional scalar field corresponds to a flat direction in the inflaton potential. Within asymptotically safe scalar-tensor theories, this feature is a direct consequence of the observation that the underlying NGFP does not come with a relevant direction which could generate a non-trivial scalar potential. This feature may change once suitable sets of fermions and Yukawa-couplings are included in the analysis \cite{Pastor-Gutierrez:2022nki} or includes a kinetial \cite{Wetterich:2022ncl}. Once the corresponding scalar potential is understood at sufficient detail, it would be very interesting to investigate whether scalaron-driven inflation is still an attractive phenomenological scenario linked to the asymptotic safety mechanism. We emphasize that while the covariant multi-field formalism developed in Sect.\ \ref{sect.5} reduces to an effective single-field description in the present context, due to the shift symmetry protecting the $\phi$-direction, this reduction is a consequence of the specific fixed-point structure of the theory rather than an assumption. It is the absence of a relevant direction at the NGFP  that enforces this flat direction. This situation may change once the matter content of the theory is extended, and the multi-field formalism developed here will then be directly applicable without modification. We hope to come back to this point in the future.
\section{Acknowledgments}
\label{sect.acknowledgments}
The authors are grateful to J.Daas for useful discussions. C.L. acknowledges support and hospitality from Radboud University during the initial stages of this work.
\bibliography{SSTwAS}

\end{document}